\documentclass{iopart}

\usepackage{graphicx}
\usepackage{cite}
\usepackage{amssymb}

\begin{document}
\title{Optical study of orbital excitations in transition-metal oxides}
\author{
R R\"{u}ckamp\dag, E Benckiser\dag,
M W Haverkort\dag,
\newline
H Roth\dag, T Lorenz\dag, A Freimuth\dag,
L Jongen\ddag, A M\"{o}ller\ddag,
\newline
G Meyer\ddag,
P Reutler\S, B B\"{u}chner$\|$, A Revcolevschi\S,
\newline
S-W Cheong\P,
C Sekar$\|$, G Krabbes$\|$,
and M Gr\"{u}ninger\dag +
}

\address{
\dag\ II. Physikalisches Institut, Universit\"{a}t zu K\"{o}ln,
Z\"{u}lpicher Str.\ 77, D-50937 K\"{o}ln, Germany}
\address{
\ddag\ Institut f\"{u}r Anorganische Chemie, Universit\"{a}t zu K\"{o}ln,
D-50937 K\"{o}ln, Germany}
\address{\S\ Laboratoire de Chimie des Solides, Universit\'e
Paris-Sud, 91405 Orsay C\'edex, France }
\address{$\|$ IFW Dresden, D-01171 Dresden, Germany}
\address{\P\ Department of Physics and Astronomy, Rutgers University,
Piscataway, New Jersey 08854 USA }
\address{+ II. Physikalisches Institut, RWTH Aachen, D-52056 Aachen, Germany}

\begin{abstract}
The orbital excitations of a series of transition-metal compounds are studied by means of
optical spectroscopy. Our aim was to identify signatures of collective orbital excitations
by comparison with experimental and theoretical results for predominantly local
crystal-field excitations. To this end, we have studied TiOCl, RTiO$_3$ (R=La, Sm, Y),
LaMnO$_3$, Y$_2$BaNiO$_5$, CaCu$_2$O$_3$, and K$_4$Cu$_4$OCl$_{10}$, ranging from early to
late transition-metal ions, from $t_{2g}$ to $e_g$ systems, and including systems in which
the exchange coupling is predominantly three-dimensional, one-dimensional or zero-dimensional.
With the exception of LaMnO$_3$, we find orbital excitations in all compounds.
We discuss the competition between orbital fluctuations (for dominant exchange coupling)
and crystal-field splitting (for dominant coupling to the lattice).
Comparison of our experimental results with configuration-interaction cluster calculations
in general yield good agreement, demonstrating that the coupling to the lattice is important
for a quantitative description of the orbital excitations in these compounds.
However, detailed theoretical predictions for the contribution of collective orbital
modes to the optical conductivity (e.g., the line shape or the polarization dependence)
are required to decide on a possible contribution of orbital fluctuations at low energies,
in particular in case of the orbital excitations at $\approx$ 0.25\,eV in RTiO$_3$.
Further calculations are called for which take into account the exchange interactions
between the orbitals and the coupling to the lattice on an equal footing.
\end{abstract}
\pacs{78.30.-j, 71.70.Ch, 71.27.+a, 75.50.Ee}
\submitto{New Journal of Physics}

\date{February 6, 2005}
\ead{grueninger@physik.rwth-aachen.de}
\maketitle

\section{Introduction}

The large variety of interesting physical phenomena observed in transition-metal
oxides results from the strong electronic correlations within the partly filled
$3d$ shell \cite{imada}. The correlated $3d$ electrons carry charge, spin and orbital
degrees of freedom, and often they are strongly coupled to the lattice. Due to the
complex interplay of these degrees of freedom \cite{nagaosatokura}, a subtle change
of a single parameter such as the bond angle or the temperature may result in a
dramatic change of the physical properties.
A prerequisite for a quantitative description of this complex interplay is a
detailed understanding of the physics of each degree of freedom separately.
Over the last years, a lot of progress was achieved regarding the properties
of, e.g., low-dimensional quantum spin systems \cite{auerbach,tsvelik}, in which the
charge degrees of freedom are frozen out due to the formation of a Mott-Hubbard insulator
at half filling.
Many fascinating phenomena have been discovered such as the absence of long-range
order even at zero temperature in one-dimensional spin liquids \cite{hohenberg,merminwagner},
the opening up of spin gaps in even-leg $S$=1/2 ladders \cite{barnes} or in the
$S$=1 Haldane chain \cite{haldane}, and novel excitations such as spinons
\cite{faddeev,andrei,karbach,arai}, triplons \cite{schmidt} or two-triplon
bound states \cite{sushkov,windt}.
The coupling of spin and charge degrees of freedom may give rise to superconductivity
in doped spin ladders \cite{dagotto92,dagottoRev,uehara} and is one of the key features in the field of high-temperature
superconductivity in the cuprates.
The interplay of the spins and the lattice has been analyzed in great detail in
the spin-Peierls system CuGeO$_3$.

Orbital and spin degrees of freedom are intimately connected with each other, and it
is tempting to speculate about novel quantum phenomena related to the orbital
degree of freedom. Some interesting examples which have been discussed recently
are orbital liquids \cite{ishihara97,feiner,khaliullin,khaliullin01}, strong
orbital fluctuations,
an orbital Peierls state \cite{khaliullinyvo,sirker,horschyvo} and
orbital waves (orbitons) \cite{saitoh,ishihara97a,khaliullinokamoto02,khaliullinokamoto03,ishiharaytio}.
Candidates for the realization of these phenomena are LaTiO$_3$ \cite{khaliullin,keimerlatio},
TiOCl \cite{seidel03,seidel04}, YVO$_3$ \cite{ulrichyvo} and LaMnO$_3$ \cite{saitoh}, respectively.
However, many of these claims are heavily debated in the literature
\cite{cwik,cracolatio,haverkortlatio,mochizuki01,mochizuki03,mochizuki03b,mochizuki04,fritschlatio,hemberger,kiyama,harris03,harris04,kikoin,pavarini,solovyev04,nagaosa,grueninger,krueger,martincarron}.
These novel phenomena are based on the exchange interactions between orbitals on adjacent sites.
A crucial assumption is that some orbitals are (nearly) degenerate. For instance, the discussion
of an orbital liquid in pseudo-cubic titanate perovskites \cite{khaliullin} starts
from a partially filled $t_{2g}$ level and assumes that the degeneracy is lifted by
quantum fluctuations based on exchange interactions.
However, the orbital degeneracy may also be lifted by the coupling to the lattice (Jahn-Teller
effect). Drawing an analogy with the spin degrees of freedom, the situation is similar to the case
of two antiferromagnetically coupled spins in an external magnetic field. If the exchange interactions
dominate, the two spins will form a singlet. However, if the external magnetic field dominates,
the spins will align parallel to the field, and the spin fluctuations are strongly suppressed.
For the orbital degrees of freedom, we are thus confronted with two central questions:
Which of the two mechanisms yields the larger energy gain ?
How strongly are the orbital fluctuations quenched by the coupling to the lattice ?
For experimental studies, the challenge clearly is to find a compound in which the
exchange interactions between the orbitals are significant, while the coupling to the
lattice and the corresponding crystal-field splitting are relatively small.
Pseudo-cubic $3d^1$ or $3d^2$ perovskites are promising candidates, since orbital quantum fluctuations
are particularly large for $t_{2g}$ electrons (threefold degeneracy) in a cubic lattice (large
frustration of the orbital interactions) \cite{khaliullin}, while the coupling to the lattice
is smaller for $t_{2g}$ electrons than for $e_g$ electrons in octahedral symmetry.

\begin{figure}[tb]
\begin{center}
\includegraphics[width=5cm, angle=0,clip]{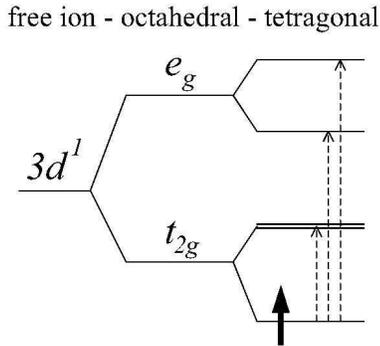}
\caption{Crystal-field splitting of the $3d$ states in an octahedral environment ($e_g$, $t_{2g}$)
and in the case of tetragonal symmetry. For a $3d^1$ ion in tetragonal symmetry, one expects
three distinct $d$-$d$ transitions.
\label{ddTrans}}
\end{center}
\end{figure}

Studies of the coupling of orbital degrees of freedom to the lattice
have a long history in the context of crystal-field transitions \cite{ballhausen}.
In a crystalline environment, the five-fold degeneracy of the $3d$ orbitals is
(partially) lifted by the electrostatic crystal field and by hybridization with the ligand ions.
For instance in local cubic symmetry one finds a splitting into a triply degenerate $t_{2g}$
level and a doubly degenerate $e_g$ level (see figure \ref{ddTrans}).
The value of this splitting is denoted by 10\,$Dq$.
For simplicity, we consider an insulating host lattice plus a Ti$^{3+}$ impurity
with a single electron in the $3d$ shell. In the ground state, this electron occupies the
energetically lowest crystal-field level. An orbital or crystal-field excitation corresponds
to the promotion of this electron into a higher-lying orbital, e.g., from the $t_{2g}$ level
into the $e_g$ level (see figure \ref{ddTrans}).
For the example of $3d^1$ Ti$^{3+}$, the excitation energy of this process typically amounts to
10\,$Dq \gtrsim$ 2\,eV \cite{kikoinbook}.
Similar values of 10\,$Dq$ are also found for other transition-metal ions.
This value is lying in the visible range and explains the vivid colors of gemstones such as
ruby (impurities of Cr$_2$O$_3$ in an Al$_2$O$_3$ host lattice).
Coming back to the analogy with the spin degrees of freedom, such a crystal-field excitation
is similar to flipping an impurity spin in a tremendous external magnetic field.
The challenge is to reduce the ``external'' field and to enter the realm where the
collective nature of the excitations becomes relevant.

In a nearly cubic environment, the splitting of the $t_{2g}$ subshell is much smaller than
10\,$Dq$. In the case of Ti$^{3+}$ impurities in an Al$_2$O$_3$ host lattice, absorption features
have been observed at about 5 and 13\,meV in infrared transmittance measurements \cite{nelson}.
The splitting was attributed to the low-symmetry crystal field and spin-orbit coupling.
However, the level splitting of 13\,meV is much smaller than predicted by crystal-field theory,
which was tentatively attributed to a reduction of the crystal field by a dynamic
Jahn-Teller effect \cite{nelson}.

Inter-site interactions between orbitals may already be important in the case
of impurities embedded into a host lattice (if the impurity concentration is large enough).
Back in 1959 it was realized that the interactions between Cr$^{3+}$ ions in concentrated
ruby give rise to new structure in the optical data \cite{schawlow}.
If the transition-metal ions form a lattice, the interactions between neighboring sites give rise
to a finite hopping probability of an excited state from one site to another. Thus the excitation
gains a finite dispersion or band width.
It can be described in terms of a Frenkel exciton, i.e., a tightly bound electron-hole pair.
The electron and the hole occupy the same site, but the pair may hop from one ion to another.
In a translationally invariant system, the exciton is described by a superposition of local
crystal-field excitations with a phase factor $\exp(\rmi k r)$, where $k$ denotes the wave vector
and $r$ runs over all transition-metal sites.
This does not play a major role in the discussion of crystal-field excitations of
$3d$ electrons in the range of 10\,$Dq$ because the dispersion usually is much smaller than both
the excitation energy and the peak width (the latter arises from the coupling to the lattice,
see the discussion of the Franck-Condon effect below).
However, experimental evidence for a finite dispersion has been derived from the analysis of
magnon-sidebands of crystal-field excitations at about 2.3\,eV \cite{sell}.

In a Mott-Hubbard insulator, the energy necessary to create a free electron-hole pair is of the order
of $U$, but the binding energy of a Frenkel exciton is also of the order of $U$. Thus the Frenkel exciton
may exist way below the band gap, and the relevant energy scale is set by the crystal-field splitting
and by the hopping amplitude of the exciton between neighboring sites.
Most noteworthy, the band width of a single doped carrier in an antiferromagnetic Mott insulator
is strongly suppressed by the coupling to the spins, so that the band width of an excitonic particle-hole
pair may even be larger than the one-particle band width \cite{hasan}.

The novel aspect of orbital waves or orbitons
\cite{saitoh,ishihara97a,khaliullinokamoto02,khaliullinokamoto03,ishiharaytio}
is that the exchange interactions are assumed to mark the dominant energy scale, minimizing
the kinetic energy. In this case, the collective nature of the excitations prevails over the
local character, in analogy to spin waves.
Note, however, that the orbital degree of freedom does not show continuous rotational symmetry.
Orbitons in an orbitally ordered state thus are gapped.
The dispersion and the excitation energy may be of comparable magnitude,
giving rise to interesting novel effects. Examples are a possible contribution to
the specific heat \cite{khaliullin,fritschlatio}, the renormalization of magnon
spectra \cite{khaliullinkilian} or the formation of orbiton-magnon bound states \cite{brink98}
due to a coupling between spin and orbital degrees of freedom.

Optical spectroscopy has proven to be an excellent tool for the study of crystal-field excitations
\cite{ballhausen,figgis}. Besides the determination of the excitation energy, it also provides
valuable information via an analysis of the line shape observed in the optical conductivity
$\sigma(\omega)$. A detailed study of orbital excitations in $\sigma(\omega)$ thus may offer
an important test whether novel collective phenomena  appear in a particular compound.
Here, we present optical data for a series of insulating transition-metal oxides
(TiOCl, RTiO$_3$ (R=La, Sm, Y), LaMnO$_3$, Y$_2$BaNiO$_5$, CaCu$_2$O$_3$, and
K$_4$Cu$_4$OCl$_{10}$),\footnote{Parts of the data on LaMnO$_3$
and on CaCu$_2$O$_3$ have been reported before in
references \cite{grueninger,advances}.}
including early and late transition-metal ions, $t_{2g}$ as well as $e_g$ systems,
and systems in which the exchange coupling is predominantly three-dimensional, one-dimensional
or zero-dimensional.
Orbital excitations are observed in the range from about 2.5\,eV down to about 0.25\,eV.\@
Only in LaMnO$_3$ we did not find any signature of orbital excitations below the band gap.
We compare the experimental data with theoretical results for the crystal-field excitations
based on the point-charge model and on configuration-interaction cluster calculations.
Our calculations satisfactorily describe all of our optical conductivity data, suggesting that the
coupling to the lattice is dominant in these compounds. However, in the case of the low-energy
modes ($\approx$ 0.25\,eV) in RTiO$_3$ we cannot exclude a contribution from orbital fluctuations.
But our results clearly demonstrate that the coupling to the lattice may not be
neglected if one aims at a quantitative description of the orbital excitations of the above
compounds. We suspect that this will turn out to be generally valid in insulating compounds.
One may hope that the importance of low-energy orbital excitations is enhanced by choosing
a system which is closer to a metal-insulator transition.

In the following section, we will discuss the selection rules for the observation of orbital
excitations by means of optical spectroscopy.
In the subsequent section \ref{sect_theo}, the point-charge model and the cluster calculations
will be explained, whereas section \ref{sect_exp} is devoted to experimental aspects.
In sections \ref{sect_data_TiOCl} to \ref{sect_data_cuprates} we will present the optical data
of TiOCl, RTiO$_3$, LaMnO$_3$, Y$_2$BaNiO$_5$, CaCu$_2$O$_3$, and K$_4$Cu$_4$OCl$_{10}$,
respectively. Some general conclusions will be given in section \ref{sect_conclusion}.

\section{Orbital excitations in optical spectroscopy}
\label{sect_ddIR}

\subsection{The case of a single ion}
\label{sect_single}

In the following paragraphs we first consider the physics of a single ion embedded
into a host lattice. The effects of interactions between the transition-metal ions
will be addressed in section \ref{sect_inter}.

The dominant contribution to the optical conductivity $\sigma(\omega)$ arises from
electric dipole transitions. The matrix element for a $d$-$d$ transition induced
by a photon is proportional to
$$\langle \psi_{final} | {\bf p} | \psi_{initial}\rangle \, .$$
The dipole operator {\bf p} has odd parity. Considering a transition-metal site with
inversion symmetry, the above matrix element vanishes due to the even parity
of the $3d$ wave functions,
$$\langle\mathrm{even} |  \mathrm{odd}  | \mathrm{even}\rangle = 0 \, .$$
Hence a mere $d$-$d$ transition is forbidden within the dipole approximation in compounds
with inversion symmetry on the transition-metal site, i.e., the $d$-$d$ transitions
do not contribute to $\sigma(\omega)$. However, there are several processes which
allow the observation of $d$-$d$ transitions, but one has to keep in mind that the
corresponding features are only weak.
In the present paper we will show examples for orbital excitations observed in $\sigma(\omega)$
which are due to
(i) the absence of inversion symmetry on the transition-metal site,
(ii) a phonon-activated mechanism,
and (iii) magnon-exciton sidebands.

A very attractive way for the observation of orbital excitations opens up if the crystal
structure does not show inversion symmetry on the transition-metal site, as e.g.\ in
TiOCl (see section \ref{sect_data_TiOCl}).
In this case, parity is not a good quantum number, so that even and odd states mix.
The amount of mixing can be estimated within the point-charge model, see section \ref{sect_theo}.
It depends on the difference in energy between the even ($3d$) and odd (e.g.\ $3p$ or $4p$)
states and on how strong the deviations from inversion symmetry are.
The small spectral weight of the orbital excitations is taken away from the dipole-allowed
absorption band, e.g., from the $3d$-$4p$ transition.
The major advantage of this structurally induced mechanism is that it allows to make clear
predictions on the polarization dependence of the orbital absorption features because
the (local) {\it symmetry} of the mixed states can be determined unambiguously within the
point-charge model. We use the room-temperature structure of TiOCl as an example (see section
\ref{sect_data_TiOCl}). For this $3d^1$ compound we find that the lowest valence orbital
predominantly shows $d_{y^2-z^2}$
character\footnote{The ligands are located approximately along the diagonals of
the $yz$ plane, hence $d_{y^2-z^2}$ denotes a state from the $t_{2g}$ subshell.}
with a small admixture of $p_z$ character. The first excited state shows pure $d_{xy}$
character, while the second excited state is mixed from $d_{yz}$ and $p_y$ states.
Therefore, a dipole transition from the ground state
to the second excited state is weakly allowed for light polarization parallel to the
$y$ axis, but not for $x$ or $z$ polarization:
\begin{equation}
\label{eq_tiocl_selection}
\langle \alpha^\prime d_{yz} + \beta^\prime p_y | y | \alpha d_{y^2-z^2} + \beta p_z \rangle \neq 0\, ,
\end{equation}
\begin{equation}
\langle \alpha^\prime d_{yz} + \beta^\prime p_y | x | \alpha d_{y^2-z^2} + \beta p_z \rangle = 0\, .
\end{equation}
Such polarization selection rules offer the possibility for a straightforward experimental test.

\begin{figure}[t]
\begin{center}
\includegraphics[width=5cm, angle=0,clip]{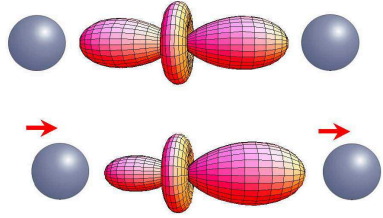}
\caption{Top: Sketch of a $3z^2$-$r^2$ orbital on a transition-metal
site in between two negatively charged ligands.
Bottom: Exciting a bond-stretching phonon breaks the inversion symmetry
on the transition-metal site, thus parity is not a good quantum number
anymore. This gives rise to a mixing of even and odd states, e.g., of
the $3d_{3z^2-r^2}$ state with the $4p_z$ state.
The sketch indicates an increased electron density on the right side,
where the distance to the negatively charged ligand has increased.
\label{3d4p} }
\end{center}
\end{figure}

If the crystal structure shows inversion symmetry on the transition-metal site,
this symmetry can be broken by an odd-symmetry phonon which is excited simultaneously
with the orbital excitation \cite{ballhausen,figgis}. This again gives rise to the admixture
of a small amount of odd character to the $3d$ wave function (see figure \ref{3d4p}).
The dependence on the polarization of the incident light is less pronounced for this
phonon-activated mechanism than for the structurally-induced mechanism described above,
because in general phonons of arbitrary polarization may contribute, i.e., $x$, $y$ and
$z$ character can all be mixed into the $3d$ states.
In order to determine the orbital excitation energy, the phonon energy has to be subtracted
from the experimentally observed peak position. One has to keep in mind that different phonons
may break the symmetry. Typically, stretching and bending modes of the metal-oxygen bonds are most
effective in doing so \cite{ballhausen}. These modes have typical energies of the order of
50-80\,meV.\@
The fact that phonons with different energies may contribute and that these phonons additionally
have some dispersion increases the width of the absorption band (the most important source
for the line width is described by the Franck-Condon effect, see below).

Another way to break the symmetry is to add impurities to the system. However, it has been shown
experimentally that this in general is by far less effective than the phonon-activated mechanism
described above \cite{ballhausen}.
One way of testing whether a phonon is involved in the infrared absorption process is to compare
the energies of the orbital excitations observed in $\sigma(\omega)$ and in Raman scattering.
In compounds with inversion symmetry, the exclusion principle states that selection rules for
Raman scattering and infrared absorption are mutually exclusive. Orbital excitations can be
observed directly in Raman scattering because {\em two} photons are involved in the scattering
process, thus the odd dipole operator has to be applied twice.
The incoming photon excites an electron from a $3d$ orbital to, e.g., a $4p$ state, from which it
falls back to an excited $3d$ state under emission of a photon.
Using again the example of a $3d^1$ system, the transition from, e.g., $d_{xy}$ to $d_{xz}$
is Raman active in crossed polarization, for instance for $y$ ($z$)
polarization of the incoming (outgoing) photon:
$$\langle d_{xz} | z | p_x \rangle\langle p_x | y | d_{xy} \rangle \neq 0 \, .$$
Other optical experiments which allow the observation of orbital excitations are, e.g.,
electroreflectance measurements \cite{falck} or third-harmonic spectroscopy \cite{dodge}.
Furthermore, $d$-$d$ excitations have been studied by means of electron energy loss spectroscopy
(EELS) \cite{fromme}.

Thus far we have neglected the spin selection rule. One has to keep in mind that optical spectroscopy
with linearly polarized light is only sensitive to spin-conserving excitations, $\Delta S$=0.
This selection rule can be relaxed by taking into account spin-orbit coupling.
Another possibility is to excite two spin-carrying modes simultaneously in such a way that the
total spin amounts to zero.
An orbital excitation from e.g.\ a triplet state to a singlet may gain a finite spectral weight
by the simultaneous excitation of a magnon, giving rise to a so-called magnon-exciton sideband
\cite{sell,tanabe,kahn}.
The spectral weight of these processes is even smaller than in the cases discussed above where the
spin was not involved. Nevertheless these processes are dominant in systems with $d^5$ ions
such as Mn$^{2+}$ \cite{sell,kahn}, in which none of the excited states carries the
same spin value as the $^6S$ ground state.
In MnF$_2$, both magnetic-dipole and electric-dipole transitions have been observed \cite{sell}.
The magnetic-dipole character can be proven experimentally by the observation of a splitting in an
applied magnetic field or by a detailed study of the polarization dependence, i.e., by showing
that the absorption features depend on the direction of the magnetic field component
and not on the electric field component.

\begin{figure}[t]
\begin{center}
\includegraphics[width=5.5cm, angle=270,clip]{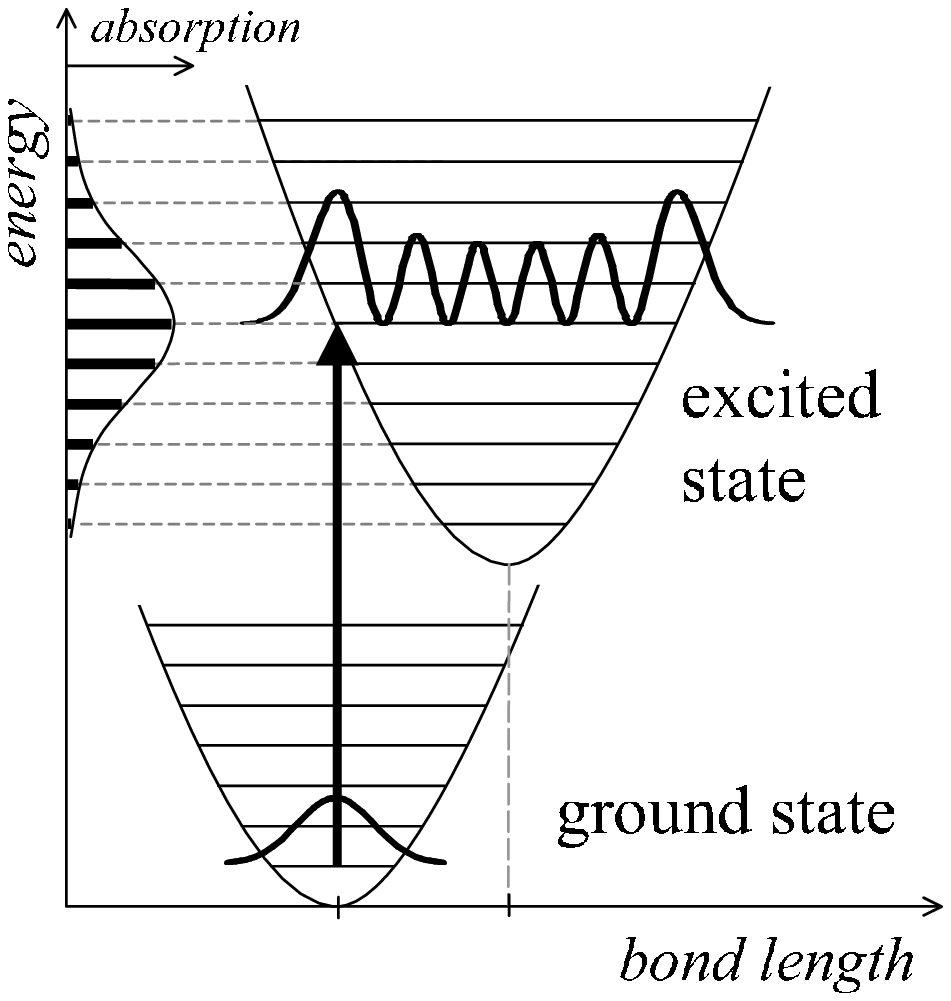}
\includegraphics[width=5.5cm, angle=270,clip]{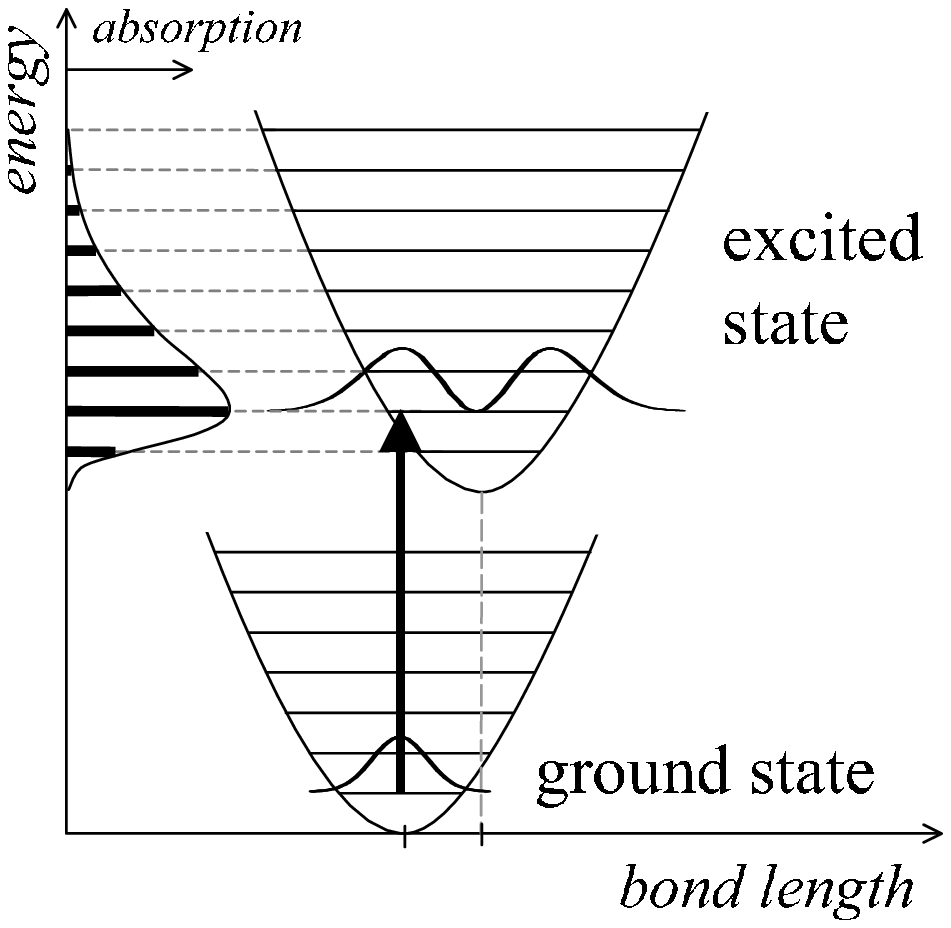}
\caption{
Sketch of the Franck-Condon principle.
In both panels, each parabola corresponds to a different orbital state and represents
the harmonic potential of the lattice. The lines within a parabola denote
phonon excitations.
The horizontal position of a parabola indicates the distance to the ligands
after the lattice has been allowed to relax in the particular orbital state.
Promoting e.g.\ an electron in an octahedral oxygen cage from $x^2$-$y^2$ to
$3z^2$ tends to push away the two negatively charged oxygen ligands on the $z$ axis.
In general, electronic time scales are much faster than the relaxation time of
the lattice. A fast electronic excitation, i.e., without relaxation of the lattice,
corresponds to a vertical transition (arrow).
The transition probability is proportional to the overlap between the
wave functions of the ground state and of the excited state.
The thick lines denote the amplitudes of the ground-state wave function and of
an excited harmonic oscillator.
The strongest overlap is obtained for the level which is closest to where
the vertical arrow cuts through a parabola.
In the final state, both the lattice and the electronic/orbital subsystem
are in an excited state, i.e., the excited states are of mixed  character
(vibrational + electronic $\rightarrow$ ``vibronic'').
Summing up the contributions from the different excited states results in
the broad absorption peak shown on the left in each panel.
Due to the dispersion of the phonons
and due to the contribution of phonons with different energies, the sharp subbands
of individual excited states are usually not resolved in a solid, yielding a single
broad band.
The width and the line shape of an absorption band in $\sigma(\omega)$ depend on the
difference in bond length of the different orbital states.
Large differences in the bond length give rise to symmetric absorption bands (left
panel), whereas small differences cause a characteristic asymmetric line shape
(right panel).
\label{frankcondon} }
\end{center}
\end{figure}

For the discussion of the line shape, one has to take the coupling to the lattice into account.
The absorption band will be broadened by phonon sidebands according to the
Franck-Condon principle, and the line shape depends on the difference of the relaxed
bond lengths of the orbital states involved (for details see the caption of figure
\ref{frankcondon}).
We emphasize that this reflects the mixed, ``vibronic'' character of the eigenmodes
(phonon + orbiton, or vibrational + electronic $\rightarrow$ ``vibronic'' \cite{ballhausen})
and thus holds irrespective of the mechanism responsible for the finite spectral weight of
an orbital excitation.
In particular, these phonon sidebands may not be confused with the phonon-activated
mechanism described above and appear also in the case of, e.g., the structurally
induced mechanism relevant for TiOCl or in Raman data.

\subsection{Interaction effects}
\label{sect_inter}

We have to address the question how to distinguish experimentally between a collective orbital excitation
and a predominantly ``local'' crystal-field excitation. A direct observation of the dispersion of the
orbital-momentum fluctuations by means
of inelastic neutron scattering would manifest a watertight proof. While neutron scattering has been
used for the study of crystal-field excitations of $f$-electron compounds \cite{guillaume},
we are not aware of such data for the case of $3d$ electrons. Here, one has to keep in mind
that the coupling to the lattice will smear out the orbital excitations significantly.
Inelastic x-ray scattering offers another $k$-dependent tool, but no collective orbital excitations
were found in the study of LaMnO$_3$, KCuF$_3$ and YTiO$_3$ presented in this focus issue \cite{tanaka}.

Under the assumption that the dominant energy scale is set by the exchange interactions,
the dispersion relations of orbitons have been calculated for the orbitally ordered states of
LaMnO$_3$, LaVO$_3$, YVO$_3$, and YTiO$_3$ \cite{saitoh,ishihara97a,ishiharaytio}, and within
a model focusing on the orbital fluctuations \cite{khaliullinokamoto02,khaliullinokamoto03}.
Predictions have been derived for inelastic neutron scattering and Raman scattering.
In section \ref{sect_single} above, we have discussed a Raman process in which the virtual
excitation into a $p$ state was assumed to take place on a single site. In case of dominant
exchange interactions, a two-site process involving the upper Hubbard band is considered
\cite{ishiharaytio}, in analogy to the well-known two-magnon Raman scattering.
For simplicity, we consider per site one electron and two orbitals.
In the first step, the incident photon promotes an electron from site 1 to site 2,
which becomes doubly occupied. In the second step, an electron hops back from site 2 to 1
under emission of a photon. In the final state, one or {\it both} electrons may be in an
excited orbital, i.e., the exchange process may give rise to one-orbiton
and/or two-orbiton excitations \cite{ishiharaytio}.
Depending on the hopping amplitudes between the different orbitals on adjacent sites,
distinct polarization selection rules have been predicted \cite{saitoh,ishiharaytio}.

The excitation of {\it two} orbitons with momenta $k_1 = -k_2$ in principle allows to probe the
orbiton dispersion throughout the entire Brillouin zone, since only the total momentum $k_1+k_2$
needs to be equal to zero in Raman scattering. Information about the dispersion is contained
in the line shape of the two-orbiton Raman band, but a detailed analysis of the line shape
encounters several problems:
(i) in general, the Raman line shape depends on the frequency of the incident photons (resonance behaviour),
(ii) the orbiton-orbiton interactions are essential for the line shape, but have not been taken
into account up to now and
(iii) the coupling to the lattice reduces the orbiton dispersion.

The optical conductivity thus far has not been considered as a tool for the investigation of orbitons.
Starting again from the crystal-field limit, we note that the optical data of a crystal-field
Frenkel exciton with a dispersion much smaller than its energy is in principle very similar to the
data of a single impurity ion embedded in a host lattice. In particular, optical spectroscopy is
restricted to the observation of excitations with momentum $k$=0, and the selection rules are the
same as for the case of a single impurity ion.
Nevertheless the dispersion may play a role if two modes are excited simultaneously, as e.g.\ in
a magnon-exciton sideband \cite{sell} or in the phonon-activated case. Only the total momentum
needs to be equal to zero, and one has to sum up contributions from excitons from the entire Brillouin
zone.

As far as the intersite exchange processes discussed above for the Raman case are concerned,
Khaliullin \cite{khaliullinpriv} has pointed out the possibility of two-orbiton-plus-phonon absorption,
similar to the two-magnon-plus-phonon absorption proposed by Lorenzana and
Sawatzky \cite{losawa95,losawa95b} for spin systems.
In systems with inversion symmetry {\it in between} adjacent sites, the exchange of two electrons
does not give rise to a dipole moment. Similar to the phonon-activated mechanism for the observation
of crystal-field transitions described above, this selection rule can be relaxed by the simultaneous
excitation of a phonon \cite{losawa95,losawa95b}.
The two-``magnon''-plus-phonon absorption\footnote{Here, we have used
the term ``magnon'' to denote spinons in spin chains, triplons in spin ladders and magnons
in a long-range ordered antiferromagnet.}
has been established as an interesting tool for studies of antiferromagnetic spin chains,
spin ladders, and layered antiferromagnets
\cite{windt,suzuura,loeder,advances,perkins93,perkins98,perkinsNi}.
Since the phonon contributes to momentum conservation, the optical conductivity probes the
two-magnon or two-orbiton spectral function throughout the entire Brillouin zone, in contrast
to two-orbiton Raman scattering, which reflects only the $k$=0 part of the two-orbiton spectrum.
Thus, both the line shape and the peak position are expected to be different in $\sigma(\omega)$
as compared to Raman data.
In spin systems, the excitation of a single magnon does not contribute to $\sigma(\omega)$ due
to the spin selection rule. In the case of orbitons, however, the phonon-activated single-site
mechanism used for the study of crystal-field excitations will also be at work if the exchange
interactions are dominant. Thus one has to expect a superposition of orbiton-plus-phonon and
two-orbiton-plus-phonon contributions.

Thus far we have discussed the two limiting cases, crystal-field excitations for dominant coupling
to the lattice and collective orbital waves for dominant exchange interactions.
Detailed theoretical predictions
for the contribution of orbital waves to the optical conductivity would certainly be very helpful
in order to distinguish experimentally between a predominantly local excitation and a collective mode.
However, a quantitative description of experimental data will require to treat both the exchange
interactions and the coupling to the lattice on an equal footing \cite{brink01}.

\section{Configuration-interaction cluster calculations and the point-charge model}
\label{sect_theo}

Thus far, no detailed predictions exist for the optical conductivity in the case of
dominant exchange interactions. We thus compare our experimental data with the
predictions for ``local'' crystal-field excitations.
Configuration-interaction (CI) cluster calculations have been performed for many years
in order to assign the correct symmetry and orbital occupancy to $d$-$d$ excitations
(see, e.g., chapter 10 of reference \cite{ballhausen}).
A typical cluster consists of the transition-metal ion and the surrounding anions,
e.g., [TiO$_6$]$^{9-}$. More distant ions are taken into account as point charges only.
The following parameters are being used: (i) the Slater integrals,
(ii) the local crystal field, and (iii) the tight-binding parameters.
\\
{\em ad (i)}
The Slater integrals describe the full local electron-electron interactions which give
rise to the main multiplet structure.
They have been obtained from Hartree-Fock calculations for a bare ion \cite{Cowan81}.
Then, these values have been reduced to 80\% in order to account for the neglect of
the $4s$ shell.
\\
{\em ad (ii)}
The crystal field or Madelung potential represents the electrostatic potential of all ions
within the crystal, which is assumed to be infinite. The ions are considered to be point
charges. The crystal field controls the on-site energies and gives rise to the energy
splitting between the orbitals.
We have calculated the crystal field using an Ewald summation, i.e., the summation is
partly performed in real space, partly in momentum space, and thus pertains to the
infinite crystal.
The orbital splitting depends on the local derivatives of the Madelung potential.
Therefore we expanded the Madelung potential in terms of spherical harmonics,
which allows to calculate the ionic crystal-field splitting if the expectation values
of $<r^k>$ are known \cite{SuganoTanabeKamimura70}, where $r$ denotes the electron coordinate
with respect to the transition-metal site and $k$ the order of the expansion.
These expectation values have been obtained from Hartree-Fock calculations \cite{Cowan81}.
\\
{\em ad (iii)}
The tight-binding parameters account for hopping processes between the ligands and
the transition-metal ion \cite{Harrison80,Slater54}. For many materials they are
well known from fits to LDA band-structure calculations. Some general rules have been
derived for the dependence of the parameters $pd\sigma$ and $pd\pi$ on the distance
between two ions \cite{Harrison80}.
Finally, the values for the on-site Coulomb repulsion on the transition-metal site ($U_{dd}$)
and on the ligands ($U_{pp}$) as well as the charge-transfer energy $\Delta$ have been
taken as reported from core-level and photo-emission spectroscopy \cite{Saitoh95}.
The cluster calculations have been performed using the code XTLS8 by A. Tanaka \cite{Tanaka94}.

In case of the $3d^1$ Ti$^{3+}$ compounds, we report both the crystal-field splitting arising
from the electrostatic potential (indicated below as {\it point-charge model}) and the
result of the cluster calculation including the hybridization with the ligands.

\section{Experimental}
\label{sect_exp}

Details concerning the crystal growth and the characterization have been described
in reference \cite{kataevtiocl} for TiOCl,
in \cite{cwik,lichtenberg} for RTiO$_3$,
in \cite{geck} for LaMnO$_3$,
in \cite{massarotti,sulewski} for Y$_2$BaNiO$_5$,
in \cite{sekar02} for CaCu$_2$O$_3$,
and \cite{deboer} for K$_{4}$Cu$_{4}$OCl$_{10}$.

Using a Fourier-transform spectrometer, we have measured the transmittance
and reflectance in the energy range from 0.01\,eV up to 3\,eV at temperatures
varying from 4\,K to 775\,K.\@
With the knowledge of both transmittance and reflectance it is possible to
directly calculate the complex optical conductivity $\sigma(\omega)$ \cite{SNS}.
For the transmittance measurements the samples were polished on both sides.
The reflectance was measured on samples with a thickness of $d \gtrapprox 1$\,mm
which were polished on only one side to prevent a contribution from multiple
reflections within the sample.
In order to determine $\sigma(\omega)$ accurately, the thickness $d$ of the
samples used for the transmittance measurement must be chosen appropriately.
Note that the transmittance depends exponentially on $d$ \cite{SNS}.
For each compound studied here, transmittance data were collected on various
samples with varying thickness $d$ (e.g., ranging from 12.5 to 300\,$\mu$m in
case of LaTiO$_3$ and from 2 to 500\,$\mu$m in LaMnO$_3$).
Thick samples ($d$ of the order of several 100\,$\mu$m)
are sensitive to weak absorption features ($\sigma < 1\,(\Omega{\rm cm})^{-1}$),
but they become opaque for larger values of $\sigma$. Thinner samples allow
to determine larger values of $\sigma$, but in the range of weak absorption
the spectra are dominated by Fabry-Perot interference fringes.
Single crystalline samples with $d < 10\,\mu$m are difficult to
handle. In particular, the surfaces of polished crystals are not absolutely
parallel to each other, so that the thickness $d$ may vary by a few $\mu$m
across the sample. This significantly complicates a quantitative analysis of
the data of very thin samples.
A reliable analysis of transmittance data measured on thin single crystals
is hence restricted to values of $\sigma(\omega)$ smaller than about
$100\,(\Omega {\rm cm})^{-1}$.
For larger values of $\sigma(\omega)$, the transmittance has to be measured on
thin films. Alternatively, a Kramers-Kronig analysis of the reflectance
or ellipsometric techniques may be used. However, measuring the transmittance
is essential for an accurate determination of weak absorption features such
as $d$-$d$ transitions.

Here, we focus on insulating samples. The accessible energy range for
transmittance measurements is therefore limited by the strong absorption
of fundamental phonon excitations below about 80\,meV and by the steep rise of
$\sigma(\omega)$ at the onset of excitations across the electronic gap.

\section{Orbital excitations in TiOCl}
\label{sect_data_TiOCl}

Recently, TiOCl has been discussed as a novel inorganic spin-Peierls system
\cite{seidel03,seidel04,kataevtiocl,beynon,imai,lemmenstiocl,caimi03,caimi04,sahadasgupta,hemberger05,shaz,abel,lemmenstiobr,sasaki,rueckamp,cracotiocl,sahadasgupta04b,pisani}.
Above about 200\,K, the magnetic susceptibility $\chi$ is described well by a model
for a one-dimensional homogeneous S=1/2 chain with exchange constant $J\! \approx \!$~676\,K
\cite{kataevtiocl}. At $T_{c1}$=67\,K, $\chi$ shows a transition to a non-magnetic
ground state \cite{seidel03,kataevtiocl}. However, indications for a second transition
(or crossover) at $T_{c2}\! \approx \!$~92\,K are observed in the magnetic susceptibility
\cite{seidel03,kataevtiocl} as well as in NMR \cite{imai} and ESR data \cite{kataevtiocl}.
Strong fluctuations above $T_{c2}\! \approx \!$~92\,K were discussed on the basis of NMR \cite{imai},
Raman and infrared data \cite{lemmenstiocl,caimi03,caimi04}.
The phase transition at $T_{c1}$ is generally interpreted as a spin-Peierls transition.
This is corroborated by the observation of a doubling of the unit cell along the
$b$ direction below $T_{c1}$ in x-ray scattering \cite{shaz,abel}.
However, the physics at higher temperatures and in particular the occurrence of a second
phase transition have not been understood so far.
It has been speculated \cite{seidel03,kataevtiocl,imai,lemmenstiocl,caimi03,caimi04,sahadasgupta,hemberger05}
that this unconventional behaviour is caused by strong orbital fluctuations,
assuming a near degeneracy of the $t_{2g}$ subshell in this distorted structure.

\begin{figure}[t]
\begin{center}
\includegraphics[width=7cm,clip]{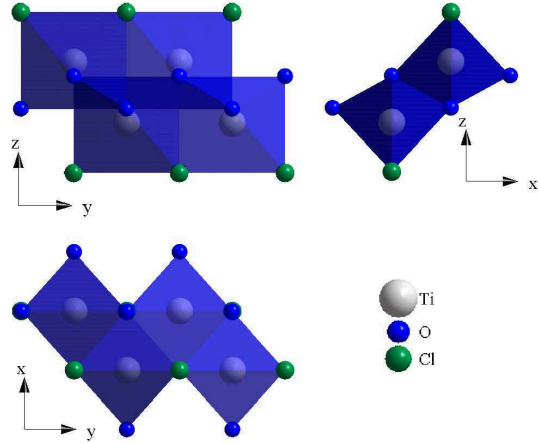}
\caption{View along different axes on four strongly distorted [TiO$_4$Cl$_2$] octahedra in TiOCl.
The $S$=1/2 chains are running parallel to the $b$ axis.
We use $x \! \parallel \! a$, $y \! \parallel \! b$, and $z \! \parallel \! c$.
For convenience, we use these axes not only at 300\,K, but also for the discussion
of the 4\,K data.
\label{tiocl_octa} }
\end{center}
\end{figure}

The structure of TiOCl consists of buckled Ti-O bilayers separated by Cl ions.
The bilayers are stacked along the $c$ direction. The [TiO$_4$Cl$_2$] octahedra
are strongly distorted (see figure \ref{tiocl_octa}). In particular, they are
strongly compressed along the $x$ axis. At room temperature, the Ti-O bond lengths
are 1.96\,\AA{} in $x$ direction, whereas the bond lengths in the $yz$ plane amount
to 2.19\,\AA{} for the Ti-O bonds and 2.40\,\AA{} for the Ti-Cl bonds \cite{snigireva}.
On each Ti site, the ground-state orbital has predominantly $y^2-z^2$ character
(see below), forming one-dimensional chains along the $b$ (or $y$) direction.
The space group is $Pmmn$ at 300\,K and $P2_1/m$ at 4\,K \cite{shaz}.
There is no inversion symmetry on the Ti site,
thus orbital excitations are directly infrared active, i.e., they contribute to
$\sigma(\omega)$ without the additional excitation of a phonon.
In this case the point-charge model does not only allow to estimate the
transition energies, but also the polarization dependence of the orbital
excitations can be predicted (see equation \ref{eq_tiocl_selection}).
In particular, strict polarization selection rules apply in the room-temperature
structure (see table 1). Below $T_{c1}$ the distortions give rise to a mixing of
the orbitals, hence the polarization selection rules are not strict anymore.
Nevertheless it is possible to give some ``effective'' selection rules because
the dipole matrix elements for the main transitions are about three orders of
magnitude (or more) larger than for the weak ones.

\begin{figure}[t]
\begin{center}
\includegraphics[width=8cm,clip]{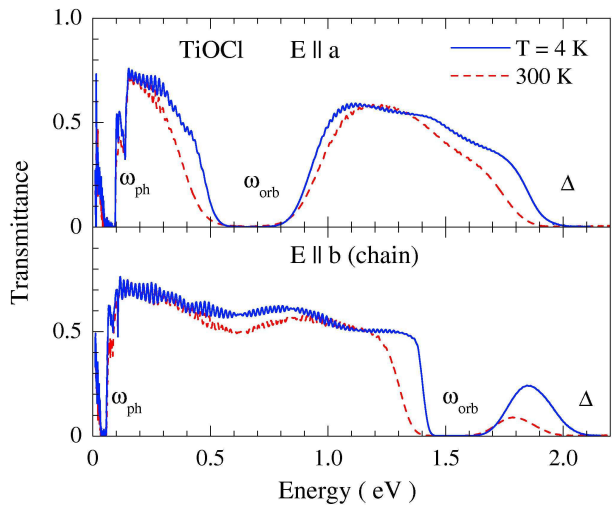}
\caption{Transmittance spectra of TiOCl with phonon absorption below $\approx$
0.1\,eV, multi-phonon peaks up to about 0.15\,eV, orbital excitations at about
0.65\,eV ($E \! \parallel \! a$) and 1.5\,eV ($E \! \parallel \! b$), and the band gap
$\Delta \approx 2$\,eV.\@
The periodic Fabry-Perot fringes in the highly transparent range indicate multiple
reflections within the sample.
\label{tiocl} }
\end{center}
\end{figure}

The transmittance measured on a thin single crystal of TiOCl is depicted
in figure \ref{tiocl} for two polarization directions, $E\! \parallel \! a$
and $E\! \parallel \! b$. Unfortunately, measurements with $E\! \parallel \! c$ could
not be performed since the available samples are very thin in the stacking direction.
Our data are in agreement with unpolarized measurements reported for energies
above 1.3\,eV \cite{maule}.
Above about 2\,eV, the sample is opaque due to excitations across the gap.
The transmittance is strongly suppressed at 0.6-0.7\,eV for $E\! \parallel \! a$
and 1.5-1.6\,eV for $E\! \parallel \! b$. The absorption feature at 0.65\,eV appears as
a weak dip also for $E\! \parallel \! b$, and the feature at 1.5\,eV gives rise
to a weak shoulder for $E\! \parallel \! a$.
An interpretation in terms of phonons or magnetic excitations can be excluded at
these energies. The excitation energies and in particular the polarization
dependence are in good agreement
with the results of the cluster calculation (see table \ref{tabletiocl}).
Thus these features can unambiguously
be attributed to orbital excitations.
As far as the polarization selection rules are concerned, the appearance
of weak features in the other polarization at 300\,K
can be attributed either to a small misalignment
of the polarizer, to absorption due to the phonon-activated mechanism, or to a small
admixture of $xy$ character to the ground state (see below), e.g. by spin-orbit coupling
or due to the dispersion (i.e., away from the $\Gamma$ point).
The absorption at about 1.5\,eV shows an asymmetric profile with a steep drop of
the transmittance on the low-energy side, in agreement with the expectations
for phonon sidebands in case of small changes of the relaxed bond length
(see figure \ref{frankcondon}). A precise determination of the line shape
requires measurements on a thinner sample, in which the transmittance is not
suppressed to zero.
The data for $T=4$\,K and for 300\,K are very similar, but the line width
increases with increasing temperature.

In the cluster calculation for TiOCl, we have to take into account the hybridization between
the Ti ions and both, O as well as Cl ions. In comparison to for instance LaTiO$_3$ or
LaMnO$_3$, we thus have two additional parameters, namely the charge-transfer energy
$\Delta_{\rm Cl}$ between Ti and Cl, and the Ti-Cl hybridization $t_{\rm Cl}$.
In fact, an accurate description of hybridization effects is essential in order to reproduce the
splitting between the $t_{2g}$ and $e_g$ subshells.
Due to the strong compression of the octahedra along $x$, the electrostatic contribution of
the point-charge model predicts that the lowest excited state has $d_{yz}$ character,
i.e., belongs to the $e_g$ subshell.\footnote{The ligands are located approximately along the
diagonals of the $yz$ plane, hence $d_{yz}$ denotes a state from the $e_g$ subshell.}
At the same time, the polarization selection rules given in table 1 predict for 300\,K
that this transition to the $d_{yz}$ level (with a small admixture of $y$ character) can be
observed for $E \! \parallel \! b$. Experimentally, the corresponding absorption feature for
$E \! \parallel \! b$ is found at about 1.5\,eV, i.e., more than 1\,eV higher than predicted by
the point-charge model. However, the energy of this transition can be described correctly by
taking into account the hybridization between Ti and its ligands, which typically adds more
than 1\,eV to the splitting between $t_{2g}$ and $e_g$ subshells.

The parameters $U_{\rm dd}$=4\,eV, $\Delta_{\rm O}$=5\,eV, and $\Delta_{\rm Cl}$=3\,eV
are estimated following the LDA+U results of reference \cite{sahadasgupta}.
The orbital excitation energies depend only weakly on $U_{\rm dd}$.
In order to model the relative strength of the Ti-O and the Ti-Cl hybridization, we have to
consider the larger ionic radius of Cl compared to O
($r_{\rm Cl}-r_{\rm O} \! \approx \!$~0.4\,\AA{})
as well as the larger polarizability.
For the former we assume that the hybridization is the
same for the two ligands if the bond length equals $r_{\rm Ti}+r_{\rm X}$ (X=O, Cl).
Additionally, the larger polarizability of Cl compared to O is modelled by a factor $t^*$ by
which the Ti-Cl hybridization is further enhanced. Good agreement between the calculated energies
and the experimental results at both temperatures is obtained for $t^* \! \approx \! 1.3$ (see
table 1).\footnote{Enhancing $t_{\rm Cl}$ by $t^*$
is necessary in order to push the $d_{yz}$ level (from the $e_g$ subshell) up to 1.5\,eV.\@
With increasing $t^*$, the energy of the $d_{yz}$ orbital increases strongly, whereas the energies
of $d_{xy}$ and $d_{xz}$ (from the $t_{2g}$ subshell) decrease.
For $t^*$=1 we find the following excitation energies: 0.32, 0.79, 1.32, and 2.21\, eV at 4\,K
for the site labeled Ti$_a$ in table 1.
}

\begin{table}[b]
\caption{
Crystal-field splitting of $3d^1$ Ti$^{3+}$ in TiOCl and polarization dependence
for infrared absorption at 300\,K and 4\,K.\@
Comparison of experimental data (see figure \ref{tiocl}) and theoretical results obtained
using the point-charge model and a cluster calculation (see section \ref{sect_theo}).
The cluster calculation uses $U_{\rm dd}$=4\,eV, $\Delta_{\rm O}$=5\,eV, $\Delta_{\rm Cl}$=3\,eV,
and an enhancement of the Ti-Cl hybridization by $t^*$=1.3 (see text).
The calculations are based on the 300\,K structure reported in reference
\cite{snigireva} and on the 4\,K data of reference \cite{shaz}.
At 4\,K, there are two inequivalent Ti sites.
For convenience, we use $x \! \parallel \! a$, $y \! \parallel \! b$, and $z \! \parallel \! c$
at both temperatures. All energies are given in eV.\@
At 300\,K, the given polarization selection rules are strict.
Due to the lower symmetry at 4\,K, only ``effective'' selection rules survive in the sense that
the dipole matrix elements for the ``main'' transitions indicated at 4\,K are about three orders
of magnitude (or more) larger than those not given in the table.
\label{tabletiocl} }
\begin{indented}
\item[]\begin{tabular}{@{}llccccc}
\br
theory: & character           & $y^2\!\!-\!z^2$  & $xy$  & $xz$     & $yz$                  & $3x^2\!\!-\!r^2$ \\
        & admixture (300\,K)  & $z$              &  --   & $x$      & $y$                   & $z$ \\
        & main admixture (4\,K) & $y,z$          &  $x$  & $x$      & $y,z$                 & $y,z$ \\
%% \mr
        & point charge (300\,K) & 0  & 0.39 & 0.68                  & 0.34                  & 1.28  \\
        & cluster (300\,K)      & 0  & 0.25 & {\bf 0.69}            & {\bf 1.24}            & 2.11  \\
       & cluster (4\,K, Ti$_a$) & 0  & 0.26 & {\bf 0.73}            & {\bf 1.53}            & 2.20  \\
       & cluster (4\,K, Ti$_b$) & 0  & 0.25 & {\bf 0.77}            & {\bf 1.47}            & 2.18  \\
        & polarization (300\,K) &    & -    & $E \! \parallel \! a$ & $E \! \parallel \! b$ & $E \! \parallel \! c$  \\
     & main polarization (4\,K) &   & $E \! \parallel \! a$ & $E \! \parallel \! a$ & $E \! \parallel \! b,c$ & $E \! \parallel \! b,c$  \\
\mr
exp.:   & energy                &    & -    & {\bf 0.65 }           & {\bf 1.5 }            & -  \\
        & polarization          &    & -    & $E \! \parallel \! a$ & $E \! \parallel \! b$ & -  \\
\br
\end{tabular}
\end{indented}
\end{table}

The energy of the lowest excited state ($xy$ orbital) is crucial in order to determine
whether orbital fluctuations are the correct explanation for the interesting physics
observed in TiOCl.
In the room-temperature structure, the transition to the first excited state is not
directly infrared active, but it becomes directly infrared active in the distorted
low-temperature structure below $T_{c1}$. According to our cluster calculation, the
lowest excited state is
expected at about 0.2-0.25\,eV.\@ However, our infrared data do not show a distinct
absorption feature in this range (see figure \ref{tiocl}).
A rough estimate of the spectral weight can be obtained from the point-charge model,
which predicts that the dipole matrix element at 4\,K is about one order of magnitude
smaller than for the transition to the $xz$ orbital. Due to a factor of $1/\omega$,
this means that the spectral weight of the excitations to the $xy$ and to the $xz$
orbital should be comparable in $\sigma(\omega)$.
One possible explanation for the lack of a corresponding feature in our infrared data
is that the first and the second excited states are nearly degenerate, as indicated
by band-structure results \cite{seidel03,sahadasgupta}.

The scenario of strong orbital fluctuations assumes a near degeneracy of the states
$y^2\!\!-\!z^2$ and $xy$. However, a sizeable admixture of the state with $xy$ character
to the ground state would have drastic consequences for the selection rules.
A transition from $xy$ to $xz+x$ ($yz+y$) will give rise to absorption for
$E\! \parallel \! b$ ($E\! \parallel \! a$), i.e., just the opposite of the
selection rules derived for the transitions from the $y^2\!\!-\!z^2$ state.
This may explain the {\em weak} features at about 0.65\,eV for $E\! \parallel \! b$
and 1.5\,eV for $E\! \parallel \! a$, but at the same time the weakness of these
features compared to the strong absorption in the perpendicular direction puts
a severe limit to the admixture of $xy$ character to the ground state.
Moreover, the $g$ factor observed in ESR spectroscopy is close to 2 \cite{kataevtiocl},
which indicates that the orbital moment is quenched by a significant splitting
($\geq 0.2$\,eV) within the $t_{2g}$ subshell.
A sizeable splitting of the $t_{2g}$ subshell is in agreement with recent LDA+U
and LDA+DMFT results \cite{cracotiocl,sahadasgupta04b,pisani}.
However, the degree of orbital polarization still needs to be clarified.
In reference \cite{cracotiocl}, the lowest orbital ($y^2\!\!-\!z^2$ in our notation)
is populated by only 70\%, indicating the possible importance of inter-orbital
fluctuations, whereas a population with 0.98 electrons was reported
in reference \cite{sahadasgupta04b}.

Both our cluster calculation and in particular the observed polarization depen\-dence
show that there is no significant admixture of the $xy$ orbital to the ground state.
Thus orbital fluctuations are clearly suppressed.
In order to understand the interesting physics of TiOCl it is therefore sufficient to
consider the interplay of lattice and spin degrees of freedom.
We suggest that the occurrence of two phase transitions results from the frustration
of interchain interactions in this peculiar bilayer structure \cite{rueckamp}.

The remarkable splitting of 0.65\,eV of the $t_{2g}$ subshell is caused by
the strong distortions of the [TiO$_4$Cl$_2$] octahedra and by the different
charges on O and Cl sites within an octahedron.
This large splitting clearly shows that $t_{2g}$ systems are not necessarily good
model compounds for the study of orbital effects based on (near) orbital degeneracy
within the ground state.

\section{Orbital excitations in RTiO$_3$ (R=La, Sm, Y)}
\label{sect_data_RTiO}

One of the novel quantum phenomena proposed in the field of orbital physics are orbital
liquids, in which long-range orbital order is suppressed by quantum fluctuations
\cite{ishihara97,feiner,khaliullin,khaliullin01}.
Based on neutron scattering results, the $3d^1$ pseudo-cubic perovskite LaTiO$_3$ has
been discussed as a realization of an orbital liquid \cite{khaliullin,keimerlatio}.
In comparison to compounds with $e_g$ electrons, $t_{2g}$ systems are promising
candidates for interesting orbital phenomena due to the threefold orbital degeneracy
and due to the smaller coupling to the lattice.
However, the results for TiOCl discussed above show that also the splitting of the
$t_{2g}$ subshell can be significant. In LaTiO$_3$, a detailed study of the structure
revealed a sizeable distortion of $\approx$ 3\%, lifting the orbital degeneracy \cite{cwik}.
At this stage, it is heavily debated whether LaTiO$_3$ represents an orbital liquid
\cite{khaliullin,keimerlatio,cwik,cracolatio,haverkortlatio,mochizuki01,mochizuki03,mochizuki03b,mochizuki04,fritschlatio,hemberger,kiyama,harris03,harris04,kikoin,pavarini,solovyev04}.
The scenario of an orbital liquid requires that the reduction of the ground-state energy
by quantum fluctuations is larger than the energy splitting of the orbitals due to the
distortion. A study of the orbital excitations and the determination of the excitation
energies are very interesting in this context.

\begin{figure}[t]
\begin{center}
\includegraphics[angle=0,width=7cm,clip]{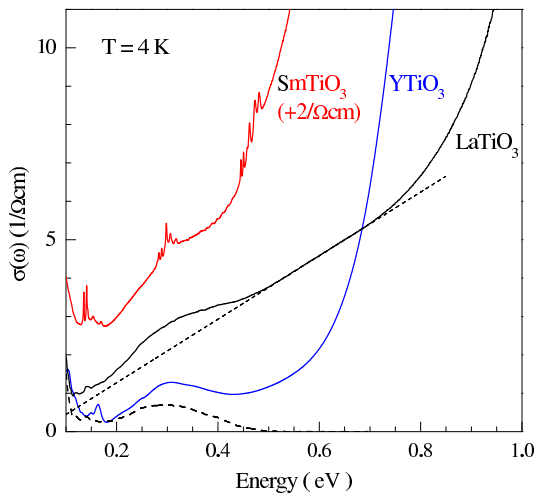}
\caption{Optical conductivity of (twinned) single crystals of LaTiO$_3$, SmTiO$_3$ and YTiO$_3$
at $T$=4\,K.\@ An offset of 2\,($\Omega$cm)$^{-1}$ has been added to the data of SmTiO$_3$
for clarity.
Phonon-activated orbital excitations are observed at 0.3\,eV in all three compounds.
For LaTiO$_3$, an estimate of the orbital excitation band has been obtained by
subtracting a linear background (dashed lines).
The additional sharp features in SmTiO$_3$ at 0.15, 0.3 and 0.45\,eV are due to
crystal-field transitions within the Sm $4f$ shell \cite{barba}.
\label{LaYSmTiO3_sig} }
\end{center}
\end{figure}

In RTiO$_3$, the magnetic ordering changes as a function of the ionic radius of the R ions
from antiferromagnetic for R=La to ferromagnetic for R=Y.\@ Within the orbital-liquid model
it has been proposed that the transition from antiferromagnetic order to ferromagnetic order
should define a quantum critical point. SmTiO$_3$ is still antiferromagnetic, but lies close
to the critical value of the ionic radius \cite{maclean}.
By measuring both the transmittance and the reflectance on twinned single crystals we were
able to observe phonon-activated orbital excitations in LaTiO$_3$, SmTiO$_3$ and YTiO$_3$
at about 0.3\,eV, i.e., in the frequency range between the phonons and the band gap (see
figure \ref{LaYSmTiO3_sig}).
In YTiO$_3$, the Hubbard gap can be identified with the onset of strong absorption at
about 0.6\,eV.\@
In LaTiO$_3$, the optical conductivity shows strong absorption above 0.8\,eV, but an
absorption tail extends down to about 0.2\,eV.\@
In a recent LDA+DMFT study, the Hubbard gap of LaTiO$_3$ (YTiO$_3$) was reported to be
0.3\,eV (1\,eV) \cite{pavarini}.
This issue of the onset of interband excitations will be discussed elsewhere.
Here, we focus on the weak absorption features at about 0.3\,eV observed in LaTiO$_3$,
SmTiO$_3$ and YTiO$_3$.

An interpretation in terms of phonons can be excluded since phonon absorption is restricted
to below $\approx$ 80\,meV.\@ The small peak in YTiO$_3$ at about 160\,meV typically marks
the upper limit for two-phonon absorption in transition-metal oxides with perovskite
structure \cite{grueninger}.
Absorption of three or more phonons has to be much weaker.
According to the magnon energies observed by inelastic neutron scattering
\cite{keimerlatio,ulrichytio}, the energy of 0.3\,eV is much too high also for
phonon-assisted magnetic absorption (i.e., two magnons plus a phonon \cite{losawa95,losawa95b}).
The very sharp additional absorption lines observed in SmTiO$_3$ at about 0.15, 0.3 and 0.45\,eV
are attributed to crystal-field transitions within the Sm $4f$ shell \cite{barba}. These lines are
much narrower than the $d$-$d$ bands because in case of the $4f$ levels both the coupling to the
lattice and the coupling between nearest-neighbor sites are much smaller.

Our interpretation of the features at about 0.3\,eV in terms of orbital excitations
is strongly corroborated by Raman scattering data of LaTiO$_3$ and YTiO$_3$. A detailed
analysis of the Raman data can be found in reference \cite{ulrichRaman}.
The Raman spectra show similar features as the
optical conductivity but shifted to lower energies by 50-70\,meV.\@
As discussed above, $d$-$d$ transitions have even parity and are thus Raman active,
whereas a contribution to $\sigma(\omega)$ arises only due to the simultaneous excitation
of a phonon breaking the inversion symmetry on the transition-metal site.
The observed shift of 50-70\,meV is in good agreement with the energies of the Ti-O bond-bending
and bond-stretching phonon modes, which are expected to yield the dominant contributions
(see section \ref{sect_ddIR}). Moreover, the transition probability for such a multi-particle
excitation (``orbiton'' plus phonon) is small, in agreement with the small absolute value of
$\sigma(\omega)$.

\begin{table}[t]
\caption{Crystal-field splitting of $3d^1$ Ti$^{3+}$ in LaTiO$_3$, SmTiO$_3$ and
YTiO$_3$ as calculated in the point-charge model and in a cluster calculation
($U_{\rm dd}=4$\,eV, $\Delta=4$\,eV).
All values are given in eV.\@
For comparison with the optical conductivity data in figure \ref{LaYSmTiO3_sig},
the energy of the symmetry-breaking phonon has to be added.
\label{tableTi} }
\begin{indented}
\item[]\begin{tabular}{@{}lllllll}
\br
La & point charge & 0 & 0.21 & 0.23 & 0.9  &  1.0  \\
   & cluster      & 0 & 0.24 & 0.26 & 2.2  &  2.4   \\
Sm & point charge & 0 & 0.15 & 0.26 & 0.9  &  1.0 \\
   & cluster      & 0 & 0.21 & 0.31 & 2.2  &  2.5 \\
Y  & point charge & 0 & 0.14 & 0.28 & 0.9  &  1.0  \\
   & cluster      & 0 & 0.19 & 0.33 & 2.2  &  2.4  \\
\br
\end{tabular}
\end{indented}
\end{table}

We thus conclude that the orbital nature of the features is unambiguous.
The orbital excitation energy amounts to about 0.2-0.25\,eV (without the symmetry-breaking
phonon). The central issue is whether these features reflect the collective nature
of orbital fluctuations \cite{khaliullinpriv} or whether they have to be interpreted as
local crystal-field excitations.
In principle, both the superexchange coupling between the orbitals and the coupling of
the orbitals to the lattice will contribute to the excitations, thus the question has
to be addressed on a quantitative level. Theoretical treatments which take into account
both kinds of coupling on an equal footing clearly would be very interesting.

At this stage, such calculations are not available. Thus we focus on a comparison of our data
with the results for local crystal-field excitations obtained within the point-charge model
and a cluster calculation (see section \ref{sect_theo}).
We have used the structural data of reference \cite{cwik} for LaTiO$_3$,
of reference \cite{maclean} for SmTiO$_3$, and of reference \cite{ytiostructure} for YTiO$_3$.
As discussed for TiOCl, the point-charge model underestimates the splitting between the
$t_{2g}$ and $e_g$ subshells by more than 1\,eV.\@ At the same time, the predictions of the
point-charge model and of the cluster calculation for the splitting of the $t_{2g}$ subshell
are rather similar, with a maximum difference of 0.06\,eV.\@
We find good agreement between theory and experiment concerning the peak energy of
0.2-0.25\,eV in all three compounds (see table \ref{tableTi}).
Similar values for the $t_{2g}$ splitting result from a recent LDA+DMFT study, in which the
covalency between R and O ions has been identified as the driving force for the
distortions \cite{pavarini}.
Significantly smaller values (27 and 181\,meV for YTiO$_3$; 54 and 93\,meV for LaTiO$_3$)
have been derived from tight-binding fits of the $t_{2g}$ band structure \cite{solovyev04}.
On the basis of x-ray absorption and spin-resolved photo-emission data of LaTiO$_3$, it has
been concluded that the splitting between the ground state and the lowest excited state
is about 0.1-0.3\,eV \cite{haverkortlatio}.
Our results for TiOCl (see above) show that 0.2-0.25\,eV is not an extraordinarily large value
for the $t_{2g}$ splitting of a $3d^1$ Ti$^{3+}$ compound.

The above mentioned change of the magnetic ordering pattern as a function of the ionic radius of
the R ions is accompanied by a change of the character of the distortions \cite{maclean}.
The radius of the Sm$^{3+}$ ions is close to the critical value \cite{maclean}.
One thus may have hoped to find a smaller crystal-field splitting in SmTiO$_3$. In contrast,
the orbital fluctuations model predicts that the orbital excitations for not too low energies
are very similar across the quantum critical point \cite{khaliullinpriv}, in agreement with
the experimental result.
Our calculations predict that the splitting between the first and the second excited states
increases from La via Sm to Y.\@ However, the energies are not as sensitive to the radius of
the R ions as one may have expected, and we consider these small differences as smaller than
the absolute uncertainty of the calculation.
Moreover, the experimental features are too broad and too close in energy to the excitations
across the gap in order to test this prediction.

Within the crystal-field scenario, the large line width can be
explained by the coupling to the lattice, i.e., in terms of the Franck-Condon effect (see section
\ref{sect_ddIR}). The additional excitation of a symmetry-breaking phonon in the infrared measurements
may give rise to an additional broadening in $\sigma(\omega)$ as compared to the Raman data.

At the present stage, the orbital excitation energy and the observed distortions \cite{cwik} can
be described satisfactorily within a local crystal-field scenario, in which the $t_{2g}$ splitting
is dominated by the coupling to the lattice. On the basis of our optical conductivity data, we did
not find any clear evidence for strong orbital fluctuations.
Also the isotropic spin-wave dispersion of LaTiO$_3$ observed in inelastic neutron scattering
\cite{keimerlatio} can be explained within a crystal-field scenario \cite{schmitz,schmitzJ}.
The small ordered moment may result from the combination of quantum fluctuations within the spin channel,
spin-orbit coupling and the small Hubbard gap. The latter gives rise to enhanced fluctuations both in the
charge and in the orbital channel and thus may contribute significantly to the reduction of the ordered
moment \cite{held}.\footnote{At the same time,
this is the central idea behind the orbital liquid scenario: the small gap gives rise
to strong orbital fluctuations which in turn reduce the spin order \cite{khaliullin}. }

Nevertheless, the uncertainty of the theoretical predictions for the crystal-field splitting is
too large to rule out a finite contribution from orbital fluctuations.
The energy of 0.2-0.25\,eV is certainly too high for a one-orbiton excitation in a collective mode
scenario. However, it possibly can be reconciled with two-orbiton excitations \cite{khaliullinpriv},
or with the sum of crystal-field and fluctuation contributions.
Additional information can be derived from the Raman data. The resonance behavior and the polarization
dependence yield evidence for a collective nature of the orbital excitations in RTiO$_3$ \cite{ulrichRaman}.
Detailed predictions for the energy and the line shape of orbitons in the optical conductivity
are necessary to further clarify this issue.

\section{Orbitons vs.\ multi-phonon peaks in LaMnO$_3$}
\label{sect_data_LaMnO}

In the manganites, the orbital degree of freedom certainly plays an important role
\cite{brinkkhomskii}.
In contrast to the threefold degeneracy of the $t_{2g}^1$ configuration of Ti$^{3+}$ ions,
the $t_{2g}^3 e_g^1$ configuration of Mn$^{3+}$ ions is doubly degenerate within
the $e_g$ orbitals in a cubic environment.
The degeneracy can be lifted by both a collective Jahn-Teller effect and by orbital
interactions.
In LaMnO$_3$, orbital order has been observed below $T_{OO}$=780\,K
\cite{murakami,rodriguez,chatterji}.
In order to explain the existence of orbital order at temperatures above the spin
ordering temperature, it is necessary to invoke the Jahn-Teller effect. It has been
proposed that the experimentally observed orbital ordering pattern can be explained
by taking into account anharmonicity \cite{brink04}.

Raman scattering data reported for orbitally ordered LaMnO$_3$ have been interpreted as
the first experimental evidence for the existence of orbitons \cite{saitoh}.
This claim is based on the observation of three Raman lines at 126, 144 and 160 meV,
on their temperature dependence and on the analysis of the polarization dependence.
Since Raman spectroscopy is restricted to $k$=0 excitations, it is not
possible to follow the dispersion of the elementary excitations. However, in the
case of LaMnO$_3$ one expects different excitation branches with different symmetries
at $k$=0, and these were identified with the three Raman lines \cite{saitoh}.
We have challenged the orbiton interpretation on the basis of a comparison with
the optical conductivity spectrum \cite{grueninger} (see also \cite{saitohreply}).

In LaMnO$_3$, the direct observation of orbital excitations is allowed in Raman spectroscopy,
but a contribution to $\sigma(\omega)$ requires to break the parity selection rule, e.g.,
via the simultaneous excitation of a phonon. Therefore, the orbital excitations are expected
to be shifted in $\sigma(\omega)$ with respect to the
Raman lines by the phonon energy of roughly 50-80\,meV (see section \ref{sect_ddIR} and
reference \cite{braden}), in agreement with the results for RTiO$_3$ discussed in the
preceding section.

\begin{figure}[t]
\begin{center}
\includegraphics[width=6cm,clip]{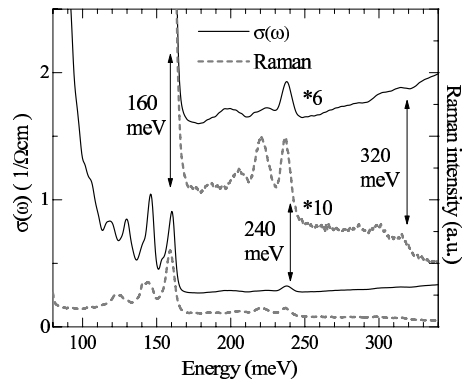}
\caption{Optical conductivity $\sigma(\omega)$ ($T$=4\,K) of LaMnO$_3$ \cite{grueninger}
showing multi-phonon features at, e.g., two, three and four times 80\,meV.\@
For comparison, we plot the Raman-scattering data ($T$=15\,K) by Choi {\it et al.} from
reference \cite{lemmenslamno}.
The two top curves focusing on the high-energy features show the same data multiplied by a factor
of 6 and 10, respectively.
\label{lamno3} }
\end{center}
\end{figure}

\begin{figure}[t]
\begin{center}
\includegraphics[width=6.8cm, angle=0,clip]{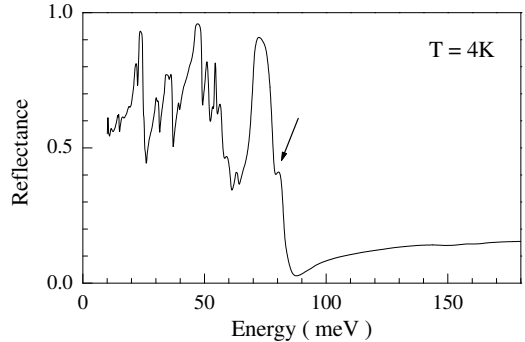}
\caption{Reflectance spectrum of LaMnO$_3$ in the phonon range.
The highest phonon mode is observed at 80\,meV, as indicated by the arrow.
\label{lamno3refl} }
\end{center}
\end{figure}

We have determined $\sigma(\omega)$ very accurately in the relevant frequency range
(see figure \ref{lamno3}) by measuring both the transmittance of thin (twinned) single
crystalline platelets and the reflectance of a sample with $d \approx 1$\,mm
(figure \ref{lamno3refl}).
The small spectral weight of the various features observed in figure \ref{lamno3} at energies
above about 80 meV, i.e., above the range of fundamental phonon absorption, is typical for
multi-phonon spectra in
insulating transition-metal oxides. For comparison, see, e.g., the spectra of $\sigma(\omega)$
of RTiO$_3$, Y$_2$BaNiO$_5$ or CaCu$_2$O$_3$ in the present work or of LaCoO$_3$ in reference
\cite{grueninger}.

In $\sigma(\omega)$ we find absorption features at about 118, 130, 146 and 160\,meV
which are very similar to the three Raman lines mentioned above.
Moreover, we identify some weak features at about 240 and 320\,meV.\@
At the latter two energies, very similar features have been observed in Raman
scattering\footnote{The Raman measurements have been performed by K.-Y. Choi, P. Lemmens
and G. G\"{u}ntherodt at the RWTH Aachen, Germany \cite{lemmenslamno}.
The Raman data have been obtained in a quasi-backscattering geometry using a laser energy
of 2.34 eV (Ar$^+$ laser).}
on the same samples by Choi and collaborators \cite{lemmenslamno} (see figure \ref{lamno3}),
using a surface that had been polished for the transmittance measurement.
The highest infrared-active fundamental phonon mode is observed in the reflectance data
at about 80\,meV (see figure \ref{lamno3refl} and references \cite{saitohreply,paolone}).
It has been interpreted as a zone-boundary Mn-O bond-stretching mode which is folded
back to $k$=0 in the orbitally ordered state \cite{saitohreply}.
A weak Raman line has been found at the same frequency \cite{martincarron,lemmenslamno}.
Given the existence of a fundamental mode at 80\,meV, the features at
160, 240 and 320\,meV are naturally explained as two-, three-
and four-phonon modes, respectively. In particular, the Raman line at 160\,meV
is certainly not too high in energy for a two-phonon mode. At 160\,meV, similar two-phonon
features are observed also in other pseudocubic perovskites such as LaCoO$_3$ \cite{grueninger}
or YTiO$_3$ (see figure \ref{LaYSmTiO3_sig}).
Three-phonon Raman scattering in LaMnO$_3$ at room temperature and above has been reported
recently in the range from 210-250\,meV \cite{martincarron}.
Multi-phonon Raman scattering is predicted to be strong in orbitally ordered LaMnO$_3$
due to the Franck-Condon effect \cite{perebeinos01} (see figure \ref{frankcondon}).

\begin{figure}[t]
\begin{center}
\includegraphics[width=6cm, angle=0,clip]{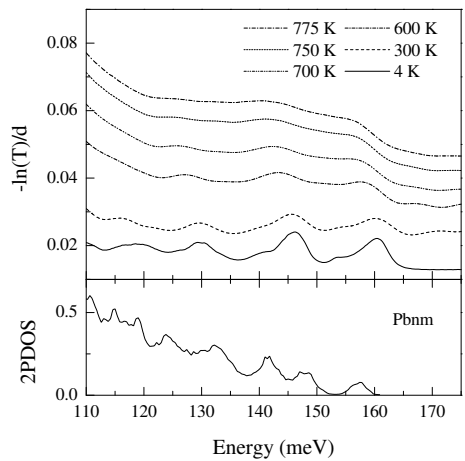}
\caption{Top: Temperature dependence of the multi-phonon features in LaMnO$_3$.
Plotting $-\ln(T)/d$, where $T$ is the transmittance and $d$ the sample thickness,
yields an estimate of the absorption coefficient \cite{SNS}.
The curves have been shifted vertically for clarity.
Bottom: The $k$=0 part of the two-phonon density of states calculated for
LaMnO$_3$ in {\it Pbnm} symmetry \cite{reichardt}.
\label{lamno3temp} }
\end{center}
\end{figure}

Let us briefly address the issue of selection rules and the problems for a theoretical
description of multi-phonon features.
The symmetry of a multi-phonon mode has to be derived using the multiplication rules of
the irreducible representations of the contributing fundamental modes. Thus an overtone of
a forbidden fundamental mode may very well be allowed. Peaks within the two- or multi-phonon
continua reflect a high density of states and do not necessarily correspond to simple multiples
of $k$=0 phonon modes.
A precise theoretical treatment of the two- and multi-phonon continua requires a detailed
knowledge of the dispersion of the fundamental modes throughout the entire Brillouin zone.
Unfortunately, such a detailed analysis of neutron scattering data has failed so far
due to the twinning of the samples \cite{reichardt}.
Moreover, multi-phonon features may depend strongly on the sample quality and on
details of the sample growth \cite{grueninger99}, which strongly complicates a
meaningful comparison with theory.
The bottom panel of figure \ref{lamno3temp} shows the $k$=0 part of the two-phonon
density of states calculated for LaMnO$_3$ in {\it Pbnm} symmetry.\footnote{The
calculations of the two-phonon density of states have been performed by
W. Reichardt (FZ Karlsruhe, Germany).}
The calculation is based on a shell model \cite{chaplot},
the parameters have been deduced from similar perovskite compounds
where the lattice dynamics was studied in detail. The highest two-phonon
peak is predicted slightly below 160\,meV, the overall structure is in
reasonable agreement with the optical data.

Within the orbiton interpretation, an explanation of features at the same energy (e.g., 160 meV)
in Raman and infrared spectroscopy requires to break the parity-selection rule without
the simultaneous excitation of a phonon. This can in principle be achieved by impurities, but the
phonon mechanism lined out above turns out to be much more effective \cite{ballhausen}.
Roughly speaking, a small impurity concentration of, e.g., 1\% breaks the selection rule
only at a small percentage of sites, whereas the phonon is effective throughout the entire
sample. This is corroborated by the shift between the Raman and the infrared data observed
in RTiO$_3$ (see above).
Moreover, the remaining differences between the Raman and the infrared spectra
-- e.g., the peak energies of 126 and 144 meV as compared to 118, 130 and 146 meV --
can easily be attributed in the multi-phonon case to the different selection rules, giving
different weight to the two-phonon density of states. However, the orbiton scenario including
impurities to account for the parity selection rule predicts {\em identical} peak energies in both
spectroscopies.

One argument favoring the orbiton interpretation was the disappearance of the relevant
Raman lines upon heating above the orbital ordering temperature $T_{OO}$=780\, K \cite{saitoh}.
We have measured the transmittance of LaMnO$_3$ at temperatures up to 775\,K (see
figure \ref{lamno3temp}), the highest temperature available in our experimental setup.
The room-temperature spectra before and after heating to 775\,K are identical, showing that the
oxygen content of the sample did not change significantly upon heating.
The absorption bands in the range of 120-160\,meV broaden strongly with
increasing temperature, but they clearly persist also at 775\.K, i.e., close to  $T_{OO}$.
The sensitivity of the Raman lines \cite{saitoh,lemmenslamno} indicates that these
multi-phonon lines are Raman forbidden in the high-temperature structure and become
Raman allowed due to the symmetry change across the phase transition.

\begin{figure}[t]
\begin{center}
\includegraphics[angle=270,width=8cm,clip]{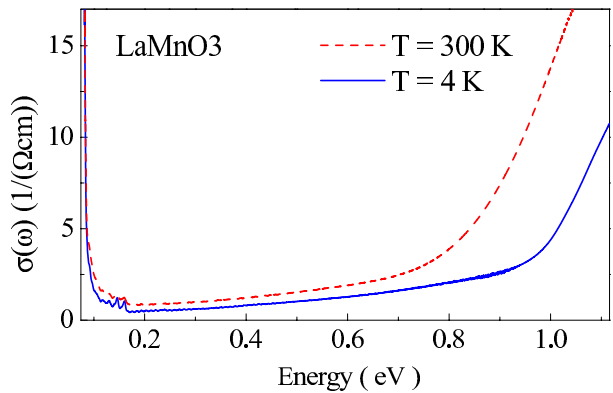}
\caption{Optical conductivity $\sigma(\omega)$ of LaMnO$_3$ in the energy range
from phonon absorption to the onset of excitations across the Mott gap.
\label{lamno3gap} }
\end{center}
\end{figure}

We have tried very carefully to find the orbital excitations at higher energies by
investigating the transmittance of several different samples of LaMnO$_3$, varying
the thickness between 2 and 500\,$\mu$m. Transmittance measurements on a thick sample
are sensitive to very weak absorption features, whereas thin samples are better suited
for the determination of larger values of $\sigma(\omega)$ (see section \ref{sect_exp}).
Our data do not show any absorption band between the multi-phonon range and the onset
of excitations across the Mott gap at about 1\,eV \cite{tobe,kovaleva} (see
figure \ref{lamno3gap}).

Allen and Perebeinos \cite{perebeinos} predicted a strong orbital absorption band centered
around 2\,eV with a Gaussian envelope of vibrational sidebands starting at about 1\,eV.\@
Additionally, they predicted a strong temperature dependence close to the onset of
absorption, in agreement with our data. A detailed comparison between theory and experiment
requires a precise determination of the line shape to still higher frequencies, i.e.,
measurements on still thinner samples.

In comparison with our data of orbital excitations observed in other transition-metal
oxides presented in this paper, the results on LaMnO$_3$ strongly suggest that the orbital
excitations are pushed to energies above 1\,eV due to the rather strong Jahn-Teller
effect in this $e_g$ electron system. Theoretical estimates of the splitting vary between
0.7 and 2\,eV \cite{perebeinos,millis,pickett,solovyev,elfimov,ahnmillis,bala}.
In a cluster calculation we have obtained a ground state quintuplet, while the two lowest
excited states are triplets lying at 1.3\,eV and 1.8\,eV, respectively
(for $U$=7.5\,eV and $\Delta$=4.5\,eV \cite{Saitoh95}).
Our calculation predicts the lowest quintuplet, which corresponds to the splitting of
the $e_g$ subshell, at about 1.9\,eV.\@
Recently, a Jahn-Teller splitting of the order of 2\,eV was derived from resonant Raman
scattering data \cite{krueger,martincarron}. A large Jahn-Teller splitting is also in
agreement with the absence of low-energy orbiton features in inelastic x-ray data \cite{tanaka}.

The eigenmodes show a mixed character (phonon-orbiton) if the coupling to the lattice
(Jahn-Teller effect) and the orbital exchange interactions are taken into account on an
equal footing \cite{brink01}.
In reference \cite{lemmenslamno}, the coupling between phonons and orbitons is discussed
on the basis of the changes observed in the (multi-)phonon Raman spectra upon variation of
temperature, symmetry or doping level.
An interpretation of the Raman features at about 160\,meV in terms of phonon-orbiton mixed
modes with predominantly orbiton character requires a rather small value for the
electron-lattice coupling \cite{brink01}.
However, if the Jahn-Teller splitting is large ($\gtrsim$ 1\,eV), the spectrum recovers the
shape predicted by the Franck-Condon effect (see figure \ref{frankcondon}) \cite{brink01}.
Note that increasing the coupling to the lattice results not only in a blue shift
of the excitation energy but also in a suppression of the orbital band width.

In summary, our search for orbital excitations below the Mott gap was not successful in LaMnO$_3$.
We are convinced that undoped LaMnO$_3$ is not a good model system for the study of
orbital waves.

\section{Y$_2$BaNiO$_5$}
\label{sect_data_YBaNiO}

\begin{figure}[t]
\begin{center}
\includegraphics[width=6cm, angle=0,clip]{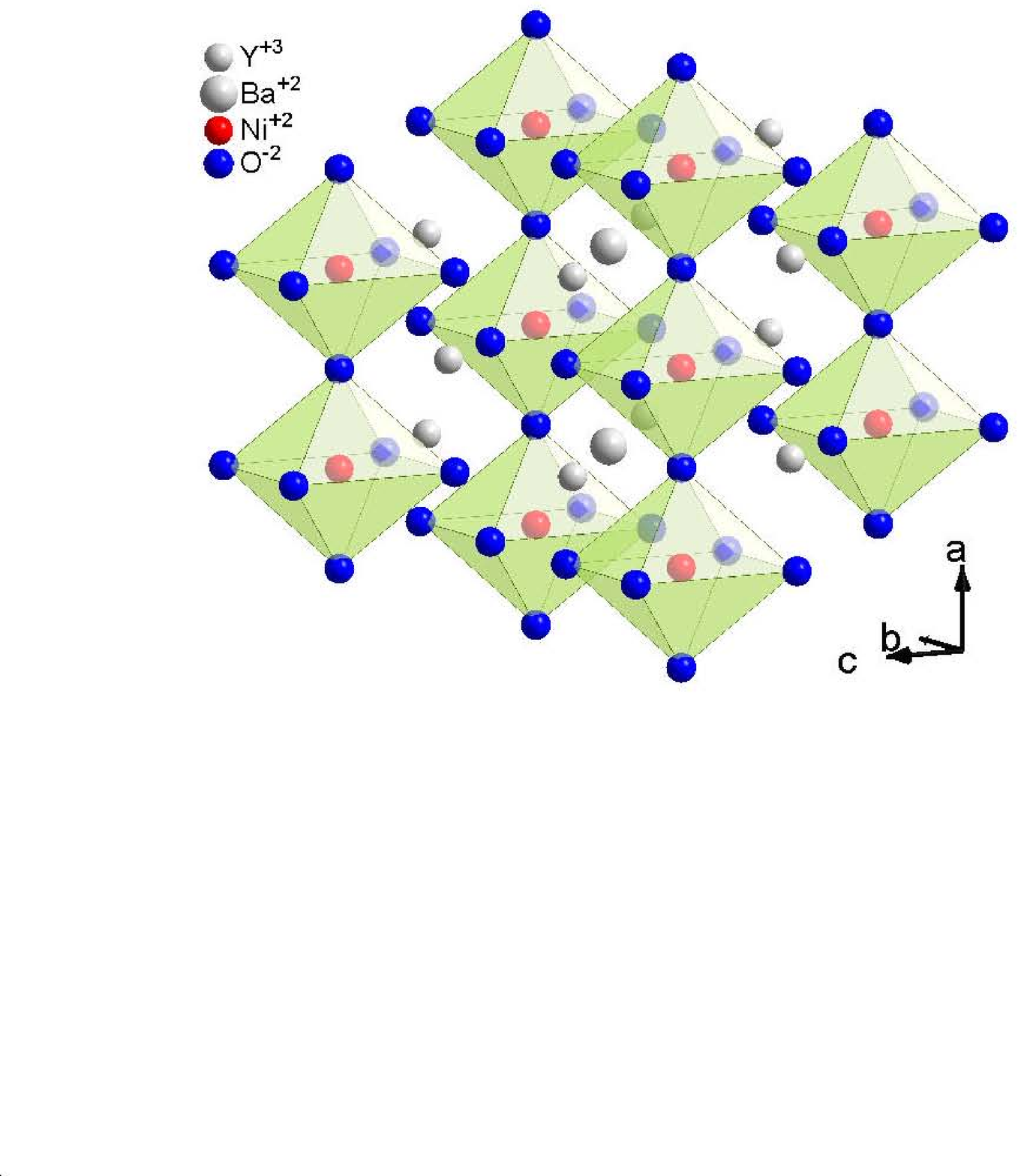}
\caption{Crystal structure of Y$_2$BaNiO$_5$ with $S$=1 chains running along the $a$ axis.
\label{ybaniostructure} }
\end{center}
\end{figure}

The compound Y$_2$BaNiO$_5$ is an example of a Haldane system, an antiferromagnetic
$S$=1 chain with an energy gap
between the collective spin-singlet ground state and the lowest excited triplet state.
The exchange coupling constant has been determined by neutron scattering,
$J \approx $ 21\,meV \cite{darriet,xu}.
The chains are formed by corner-sharing NiO$_6$ octahedra (see figure \ref{ybaniostructure})
and run along the $a$ axis. The Ni-O distance is only 1.89\,\AA{} parallel to the chains,
whereas it amounts to 2.19\,\AA{} perpendicular to the chains. Thus the octahedra are strongly
compressed.

\begin{figure}[t]
\begin{center}
\includegraphics[width=8cm, angle=0,clip]{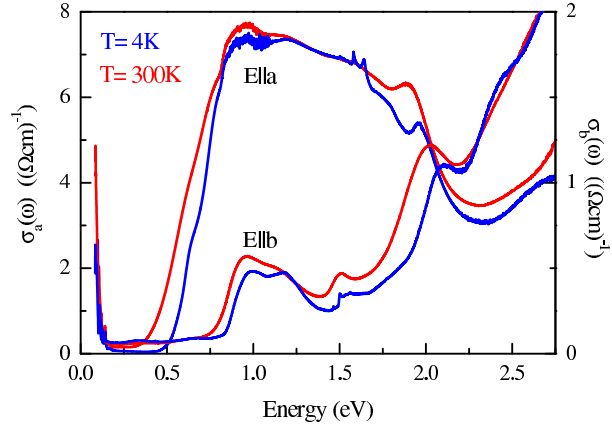}
\caption{Optical conductivity of Y$_2$BaNiO$_5$ for polarization of the electric field
vector parallel to the $a$ axis (left scale) and to the $b$ axis (right scale), respectively.
\label{ybaniodata} }
\end{center}
\end{figure}

\begin{figure}[t]
\begin{center}
\includegraphics[angle=0,width=8cm, angle=0,clip]{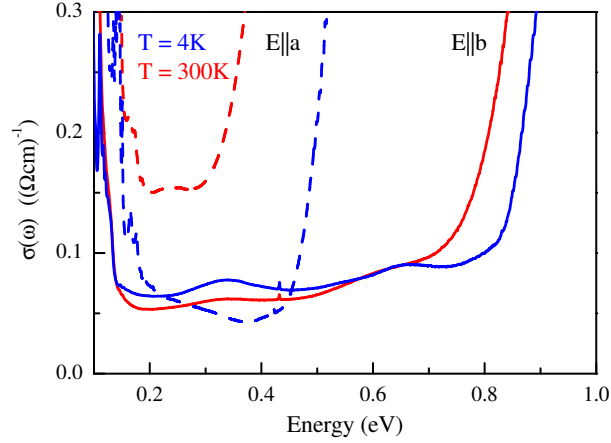}
\caption{Optical conductivity of Y$_2$BaNiO$_5$ at low energies on an enlarged scale
(cf.\ figure \ref{ybaniodata}).
\label{ybaniodatalow} }
\end{center}
\end{figure}

The optical conductivity as determined from transmittance and reflectance measurements is shown
in figure \ref{ybaniodata} for two polarization directions, $E\!\parallel \!a$ and
$E\! \parallel \! b$. In figure \ref{ybaniodatalow} we plot the spectra below 1\,eV on an
enlarged scale.
Several broad peaks are observed between 0.34 and $\approx$ 2.5\,eV in both polarization directions,
and the spectra display a pronounced polarization dependence.
As in the other compounds discussed above, phonons are limited to energies below about 80\,meV.\@
A magnetic origin can also be ruled out, since phonon-activated magnetic excitations are expected to
peak below 0.2\,eV for a $S$=1 chain with $J \approx $ 21\,meV \cite{nunnerpriv} (including a phonon
contribution of 80\,meV). We did not succeed in separating a possible phonon-activated magnetic
contribution from the multi-phonon continuum in this frequency range.
The transparency of the crystals indicates that the electronic excitations have a large gap.
The features between 0.34 and 2.5\,eV can thus be attributed to orbital excitations.
The rather complex structure of the spectra reflect the large number of multiplets of the
$3d^8$ Ni$^{2+}$ ions in this compound.

\begin{figure}[t]
\begin{center}
\includegraphics[angle=0,width=8cm,clip]{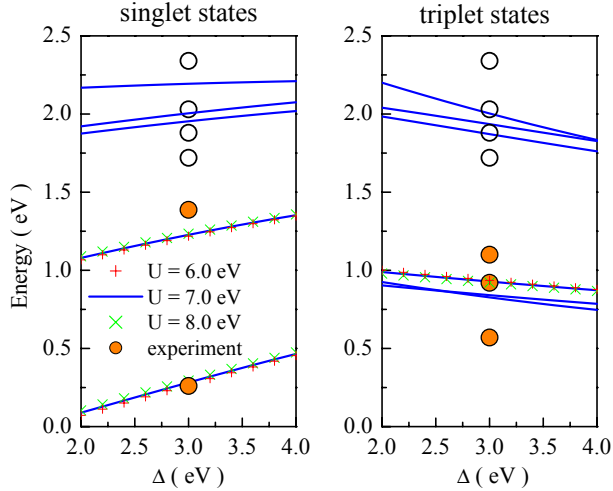}
\caption{Crystal-field levels of the $3d^8$ Ni$^{2+}$ ions in Y$_2$BaNiO$_5$
as a function of $\Delta$ as obtained in a cluster calculation.
The dependence on $U_{dd}$ (6-8 eV) is shown for a few selected levels.
The splitting of the triplet states by spin-orbit coupling is not shown for clarity.
Circles denote the experimentally determined peak positions (cf.\ figures \ref{ybaniodata}
and \ref{ybaniodatalow}). For comparison between experiment and theory, the phonon energy of
50-80\,meV still has to be added to the calculated values.
Below 1.5\,eV, the singlet/triplet character of the experimental peaks has been identified
(full circles). Above 1.5\,eV, the energy levels are too close for an unambiguous
identification (open circles).
\label{ybaniosingtrip} }
\end{center}
\end{figure}

In order to confirm this interpretation we calculated the crystal-field levels using a cluster calculation.
Here we used $U_{dd}$=7.0\,eV, $U_{pp}$=5.0\,eV, and $\Delta$=3\,eV \cite{Saitoh95,Tanaka94}.
The effect of varying $U_{dd}$ between 6 and 8\,eV and $\Delta$ between 2 and 4\,eV
is shown for a few selected levels in figure \ref{ybaniosingtrip}.
The energy levels hardly depend on the precise value of $U_{dd}$, whereas the influence of
$\Delta$ is more significant. All spin singlet energies rise with increasing $\Delta$ whereas
the spin triplet energies decrease.
For $U_{dd}$=7.0\,eV and $\Delta$=3\,eV, the cluster calculation predicts spin singlet states at
0.28, 1.23, 1.95, 2.00, and 2.19\,eV, and spin triplet states at 0.83, 0.84, 0.93, 1.87, 1.94, and
2.00\,eV, respectively. Here, we have neglected the splitting of the triplets ($\lessapprox$ 50\,meV)
by spin-orbit coupling for simplicity. Since the ground state has triplet character, the
transitions to the singlet states are forbidden by the spin selection rule $\Delta S$=0.
Comparing with the experimental data, one can identify the lowest two singlet states with the
very weak absorption band at 0.34\,eV ($E\!\parallel \!b$, see figure \ref{ybaniodatalow})
and with the sharp features at about 1.5\,eV.\@ Note that the phonon energy of 50-80\,meV still
has to be added to the calculated values.
The finite absorption strength of the feature at 0.34\,eV is attributed to spin-orbit coupling,
the sharp peaks at 1.5\,eV are addressed below.
The three peaks at about 0.65, 1.0 and 1.2\,eV are assigned to the lowest three triplet states.
The overall agreement between the calculated energy levels and the observed peak frequencies is
reasonable.

In the following, we want to focus on the two lowest spin singlet states.
The first excited singlet state at 0.34\,eV may serve as an example for a low-energy orbital
excitation in a distorted crystal structure. Its low energy arises from competing interactions.
On the one hand, the Coulomb interactions prefer a parallel alignment of the spins of the
two holes in the $3d$ shell (Hund's rule), which forces the two holes to occupy different
orbitals in the ground state (predominantly the $3z^2\!-\!r^2$ and the $xy$ orbitals, which
both point towards the negatively charged O ligands in this compound).
On the other hand, the crystal field and the Ni-O hybridization give rise to a large splitting
of these two orbitals due to the strong compression of the octahedra, favoring the $3z^2\!-\!r^2$
orbital (in the present case, the dominant contribution arises from the hybridization).
Therefore, the singlet state with both holes predominantly occupying the $3z^2\!-\!r^2$ orbital
is not much higher in energy than the spin triplet ground state. Increasing the
Ni-O hybridization by applying external pressure may lead to a further reduction of the
singlet energy.

\begin{figure}[t]
\begin{center}
\includegraphics[angle=0,width=7cm, angle=0,clip]{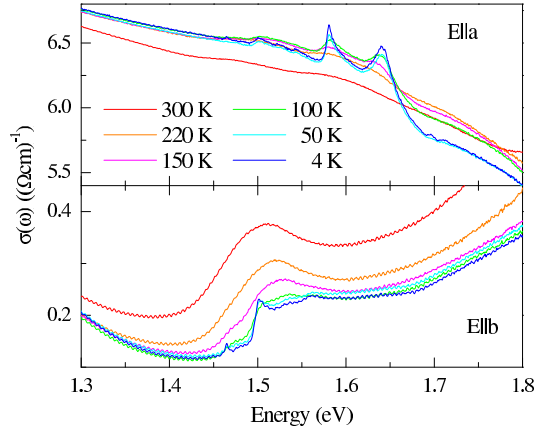}
\caption{Temperature dependence of the transition to the second excited spin-singlet
state in Y$_2$BaNiO$_5$.
The sinusoidal structure is due to Fabry-Perot interference fringes
in the transmittance data (cf figure 5).
\label{ybanio_sharp} }
\end{center}
\end{figure}

The transition to the second singlet state at about 1.5\,eV can roughly be understood as a pure
spin-flip transition, i.e., this state shows nearly the same orbital occupancy as the triplet
ground state. This explains the very narrow line width compared to all other absorption bands,
since the lattice hardly reacts to the spin flip (see the discussion of the Franck-Condon principle
in figure \ref{frankcondon}).
The fine structure is attributed to magnetic dipole transitions and to magnon-assisted electric
dipole transitions \cite{sell,tanabe,kahn}. A more detailed polarization analysis with
additional measurements on a different sample surface is required in order to determine whether
the absorption features depend on the direction of the magnetic field component
or on the electric field component.
The temperature dependence of the sharp features at about 1.5\,eV is shown in figure \ref{ybanio_sharp}.
The fine structure is rapidly washed out with increasing temperature, which is attributed to the
loss of coherence of the magnetic excitations.

\section{CaCu$_2$O$_3$ and K$_4$Cu$_4$OCl$_{10}$}
\label{sect_data_cuprates}

These two cuprate compounds show interesting magnetic properties. For the purpose of the
present paper, they serve as examples for crystal-field excitations at high energies.
In the undoped parent compounds of the high-$T_c$ superconducting cuprates, predictions
for the lowest crystal-field excitation ($x^2\!-\! y^2 \rightarrow 3z^2$) varied between
0.6\,eV \cite{grant92} and
1.9\,eV \cite{martin93} in semi-empirical and {\it ab initio} cluster calculations
\cite{grant92,martin93,eskes90,graafthesis}. Experimental evidence supporting low-lying $d$-$d$
excitations was reported on the basis of optical conductivity data \cite{geserich88,perkins93}
with a possible relevance for the mechanism of high-$T_c$ superconductivity.
However, the corresponding features at about 0.3-0.5\,eV in $\sigma(\omega)$ have
been interpreted successfully in terms of phonon-assisted magnetic absorption by
Lorenzana and Sawatzky \cite{losawa95,losawa95b}.
By means of inelastic x-ray scattering \cite{kuiper} and optical third-harmonic
generation \cite{dodge}, evidence for a splitting between the $x^2$-$y^2$ ground state
and the $3z^2$ orbital of 1.6-2\,eV has been derived.

\begin{figure}[t]
\begin{center}
\includegraphics[width=6cm, angle=0,clip]{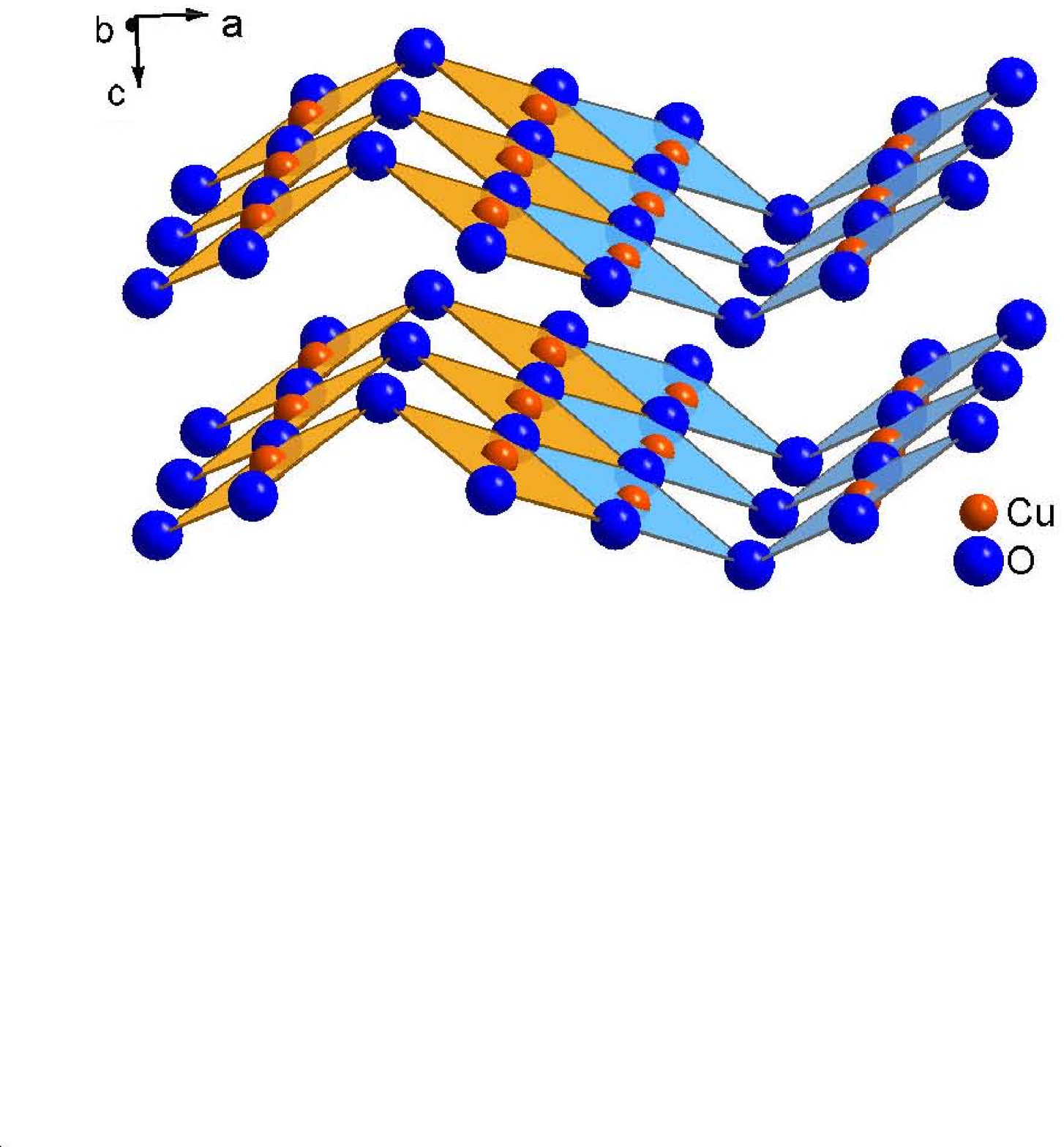}
\caption{Crystal structure of CaCu$_2$O$_3$ \cite{teske}.
\label{cacuostructure} }
\end{center}
\end{figure}

The crystal structure of CaCu$_2$O$_3$ shows buckled Cu$_2$O$_3$ layers which are stacked along the
$c$ direction \cite{teske} (see figure \ref{cacuostructure}).
Magnetically, the system is strongly anisotropic, with a dominant antiferromagnetic coupling
between the Cu spins parallel to the $b$ axis, and weak couplings in the perpendicular directions
($J_b \gg J_a \approx J_c$) \cite{advances,kimcacu2o3}.
The weak 3D coupling gives rise to long-range order below $T_N \approx $ 25\,K.\@
The inter-layer distance amounts to 3.7\,\AA{}, and the local environment of the $3d^9$ Cu$^{2+}$
ions is close to a square of four O ions. The local symmetry on the Cu site deviates only
slightly from $D_4$ symmetry. In $D_4$ symmetry, the hole occupies the $x^2$-$y^2$ orbital,
and the $xy$ orbital constitutes the first excited state, whereas the $3z^2$ orbital shows the
highest energy of the $3d$ orbitals.
In a cluster calculation we find the first excited state at 1.3\,eV.

\begin{figure}[t]
\begin{center}
\includegraphics[width=8cm, angle=0,clip]{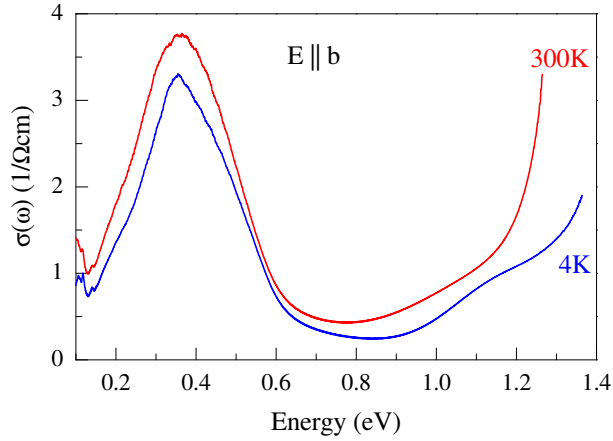}
\caption{\label{cacuodata}Optical conductivity of CaCu$_2$O$_3$, showing
phonon-assisted magnetic excitations around 0.4\,eV \cite{advances},
the charge-transfer gap at about 1.3\,eV, and a $d$-$d$ exciton just
below the gap. }
\end{center}
\end{figure}

We have determined the optical conductivity from reflectance and transmittance data
(see figure \ref{cacuodata}). Lorenzana-Sawatzky type phonon-assisted magnetic excitations
were observed at about 0.4\,eV \cite{advances}. The onset of excitations across the
charge-transfer gap is observed at about 1.2-1.3\,eV.\@
The feature at 1.1\,eV is attributed to a crystal-field excitation.
At 300\,K, both the onset of charge-transfer excitations and the crystal-field exciton
peak are smeared out significantly.

\begin{figure}[t]
\begin{center}
\includegraphics[angle=270,width=8cm,clip]{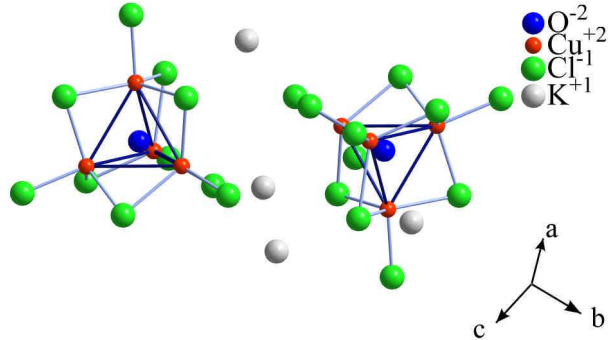}
\caption{\label{kcuoclstructure}Crystal structure of K$_4$Cu$_4$OCl$_{10}$ \cite{deboer}. }
\end{center}
\end{figure}

\begin{figure}[t]
\begin{center}
\includegraphics[width=3cm, angle=0,clip]{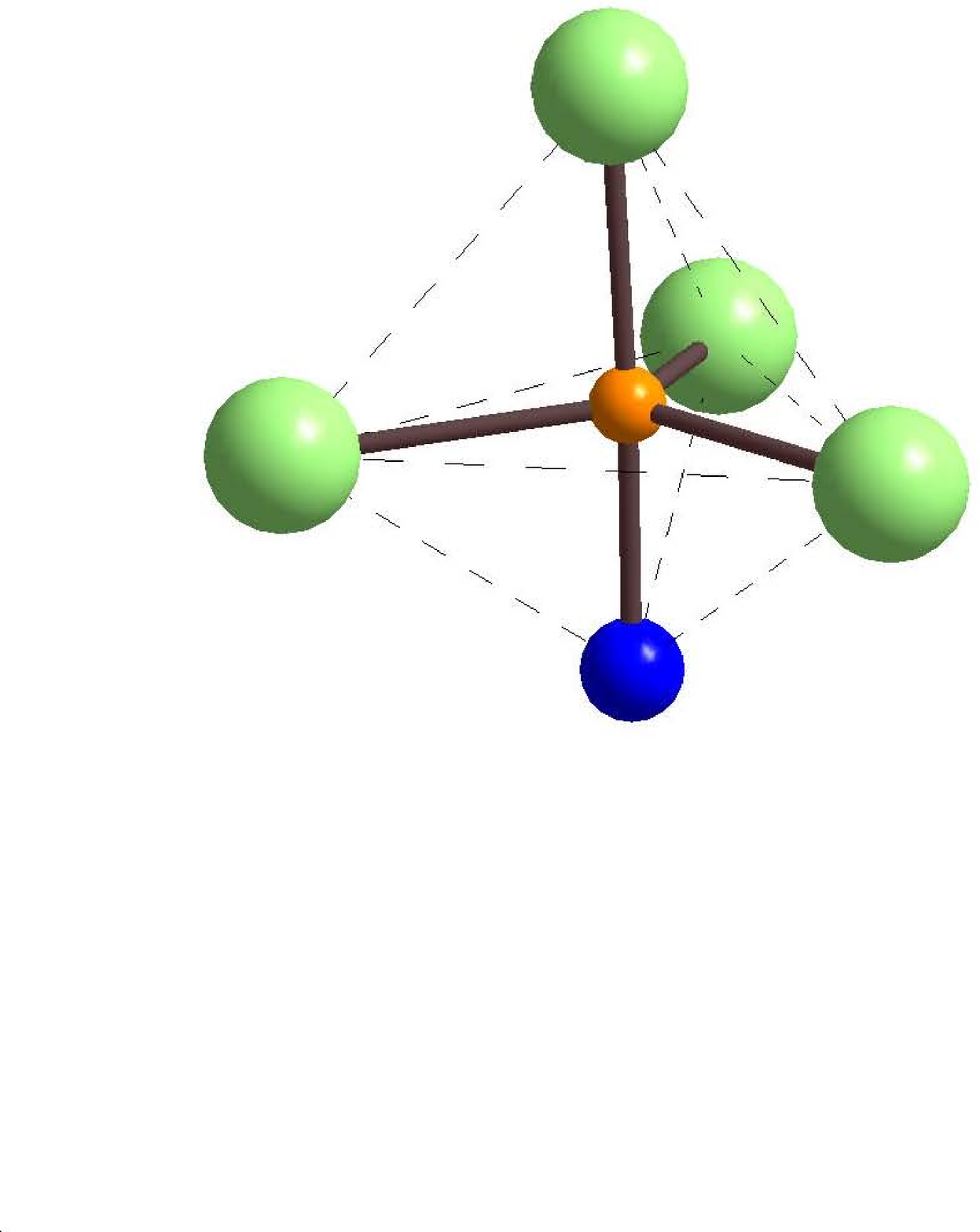}
\caption{\label{kcuoclcluster}The CuOCl$_4$ cluster considered for the calculation of the
crystal-field splitting in K$_4$Cu$_4$OCl$_{10}$. }
\end{center}
%\end{figure}
%
%\begin{figure}[t]
\begin{center}
\includegraphics[width=9cm, angle=0,clip]{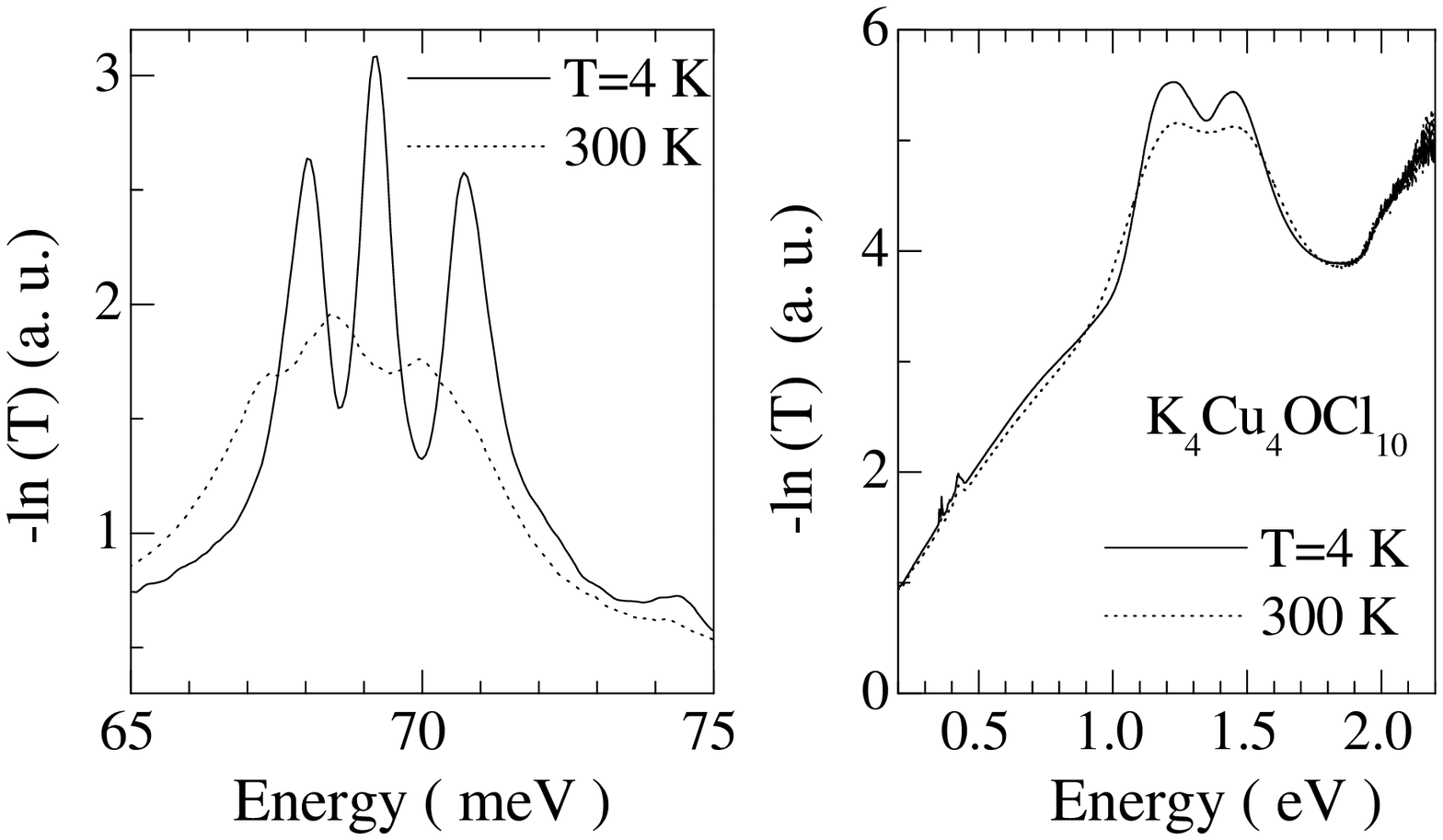}
\caption{\label{kcuocldata}Negative logarithm of the transmittance of a poly-crystalline
sample pressed in KBr.
Left: Splitting of the Cu-O bond-stretching phonon.
Right: Crystal-field excitations. }
\end{center}
\end{figure}

The compound K$_4$Cu$_4$OCl$_{10}$ has been found in volcanic sedimentations.
The crystal structure is shown in figure \ref{kcuoclstructure} \cite{deboer}.
It shows Cu$_4$ tetrahedra with a single O ion in the center of each tetrahedron
(see figure 18). Each Cu$^{2+}$ ion carries a spin of $S$=1/2 and
is surrounded by a [Cl$_4$O] cage (see figure \ref{kcuoclcluster}).
The tetrahedra are well separated by K and Cl ions. This compound thus represents
in good approximation a zero-dimensional model system for the study of local
crystal-field excitations.

As far as the magnetism is concerned, one expects two degenerate spin singlet states for
a regular, undistorted $S$=1/2 tetrahedron. The degeneracy can be lifted by magnetic
quantum fluctuations between coupled tetrahedra. The dynamics of such geometrically
frustrated quantum spin systems with low-lying singlet excitations has attracted
considerable interest \cite{yamashita,kotovtetra,lemmenscu2te2,brenigcu2te2}.
However, the degeneracy can also be lifted by distortions of the tetrahedra,
similar to the competition between orbital quantum fluctuations and lattice distortions
described above.


We have measured the transmittance $T$ of a poly-crystalline sample that was pressed
in a KBr pellet because K$_4$Cu$_4$OCl$_{10}$ is sensitive to moisture.
Plotting $-\ln(T)$ yields an estimate of the absorption, as stated above \cite{SNS}.
The Cu-O bond-stretching mode is split into three distinct absorption peaks (see
left panel of figure \ref{kcuocldata}), indicating a distortion of the tetrahedra and a
suppression of the magnetic quantum fluctuations.

Crystal-field excitations have been observed at 1.22 and 1.47\,eV (see figure \ref{kcuocldata}).
The background is attributed to scattering on grain boundaries which increases with increasing
energy.
For an estimate of the crystal-field splitting we consider a CuOCl$_4$ cluster as shown in
figure \ref{kcuoclcluster}.
Using the same parameters as for TiOCl (in particular $t^* = 1.3$,
see above), we obtain four excited states at 1.3, 1.5, 1.8 and 1.9 eV
above the ground state. A further increase of $t^*$ hardly affects
the two lower energies, but pushes the two higher levels beyond 2 eV,
in reasonable agreement with the experimental result.

\section{Conclusion}
\label{sect_conclusion}

Orbital excitations have been observed in the optical conductivity of a series of transition-metal
compounds. The orbital excitations gain a finite spectral weight in $\sigma(\omega)$ due to either
the absence of inversion symmetry on the transition-metal site, a phonon-activated mechanism,
or magnon-exciton sidebands.
In general, reasonable agreement has been obtained between the observed peak energies
and the predictions of a cluster calculation.
In the case of TiOCl, also the experimentally observed polarization dependence can be understood very
well within the point-charge model. Here, the lack of inversion symmetry on the Ti site allows to make
strict predictions for the polarization dependence in the room-temperature structure, i.e., each
transition is expected only in one particular polarization.
A pronounced polarization dependence has also been observed in case of the phonon-activated absorption
bands of Y$_2$BaNiO$_5$. However, most of these bands are observed both for $E\!\parallel \! a$ and
$E\!\parallel \! b$, i.e., only the intensity is changing as a function of the polarization.

At the present stage, our optical conductivity data do not yield clear evidence for a
collective nature of the orbital excitations in any of the compounds studied here.
The energetically lowest orbital excitations were observed in Y$_2$BaNiO$_5$ (0.34\,eV)
and in RTiO$_3$ (0.3\,eV). In Y$_2$BaNiO$_5$, the low energy can be explained by the
competition between Hund's rule coupling (favoring a spin triplet state, in which the two holes
on a $3d^8$ Ni$^{2+}$ ion occupy different orbitals) and Ni-O hybridization (favoring a
spin singlet state with both holes in the $3z^2$ orbital due to the strong compression of the
octahedra).
In the case of RTiO$_3$, detailed theoretical predictions for $\sigma(\omega)$ concerning, e.g.,
the peak energy and the line shape for absorption of collective orbital fluctuations are called
for. In particular, a quantitative estimate of the possible importance of orbital fluctuations
can only be derived from our data if the orbital exchange interactions and the coupling to the
lattice are treated on an equal footing.
However, the resonance behavior and the polarization dependence of the Raman data \cite{ulrichRaman}
give evidence for a collective nature of the orbital excitations in RTiO$_3$.
In TiOCl, on the other hand, the $t_{2g}$ splitting was found to be about 0.65\,eV.\@
A scenario of strong orbital fluctuations with a significant admixture of the $xy$ orbital
to the ground state is in contradiction with the observed polarization dependence.
We thus consider a dominant role of orbital fluctuations in this compound as very unlikely.

%%%%%%%%%%%%%%%%%%%%%%%%%%%%%%%%%%%%%55
%% acknowledgement section
%%%%%%%%%%%%%%%%%%%%%%%%%%%%%%%%%%%%%55
\ack
It is a great pleasure to acknowledge the fruitful collaborations with
C. Ulrich, B. Keimer, K.-Y. Choi, P. Lemmens and G. G\"{u}ntherodt regarding
the Raman measurements on RTiO$_3$ and LaMnO$_3$.
We are indebted to W. Reichardt for provision of the
calculated two-phonon density of states of LaMnO$_3$.
We would like to thank A. Tanaka for the use of his program XTLS8,
and A. G\"{o}ssling, D.I. Khomskii, E. M\"{u}ller-Hartmann, L.H. Tjeng, G.S. Uhrig and in
particular G. Khaliullin for many stimulating discussions.
This project is supported by the DFG via SFB 608.

\section*{References}
\begin{thebibliography}{99}

\bibitem{imada}
Imada M, Fujimori A, and Tokura Y 1998 {\it Rev.\ Mod.\ Phys.\ } {\bf 70} 1039

\bibitem{nagaosatokura}
Tokura Y and Nagaosa N 2000 {\it Science} {\bf 288} 462

\bibitem{auerbach}
Auerbach A 1994 {\it Interacting Electrons and Quantum Magnetism}, Springer,
New York

\bibitem{tsvelik}
Gogolin A O, Nersesyan A A and Tsvelik A M 1998
{\it Bosonization and Strongly Correlated Systems},
Cambridge University Press, Cambridge

\bibitem{hohenberg}
Hohenberg P C 1967 {\it Phys.\ Rev.\ } {\bf 158} 383

\bibitem{merminwagner}
Mermin N D and Wagner H 1966 {\it Phys.\ Rev.\ Lett.\ } {\bf 22} 1133

\bibitem{barnes}
Barnes T, Dagotto E, Riera J and Swanson E S 1993
{\it Phys.\ Rev.\ } B {\bf 47} 3196

\bibitem{haldane}
Haldane F D M 1983 {\it Phys.\ Rev.\ Lett.\ } {\bf 50} 1153

\bibitem{faddeev}
Faddeev L D and Takhtajan L A 1981 {\it Phys.\ Lett.\ } {\bf 85A} 375

\bibitem{andrei}
Andrei N and Lowenstein J H 1979 {\it Phys.\ Rev.\ Lett.\ } {\bf 43} 1698

\bibitem{karbach}
Karbach M, M\"{u}ller G, Bougourzi A H, Fledderjohann A and M\"{u}tter K-H 1997
{\it Phys.\ Rev.\ } B {\bf 55} 12510

\bibitem{arai}
Arai M, Fujita M, Motokawa M, Akimitsu J and Bennington S M 1996
{\it Phys.\ Rev.\ Lett.\ } {\bf 77} 3649

\bibitem{schmidt}
Schmidt K P and Uhrig G S 2003 {\it Phys.\ Rev.\ Lett.\ } {\bf 90} 227204

\bibitem{sushkov}
Sushkov O P and Kotov V N 1998 {\it Phys.\ Rev.\ Lett.\ } {\bf 81} 1941

\bibitem{windt}
Windt M, Gr\"{u}ninger M, Nunner T, Knetter C, Schmidt K P, Uhrig G S,
Kopp T, Freimuth A, Ammerahl U, B\"{u}chner B, and Revcolevschi A 2001
{\it Phys.\ Rev.\ Lett.\ } {\bf 87} 127002

\bibitem{dagotto92}
Dagotto E, Riera J and Scalapino D 1992 {\it Phys.\ Rev.\ } B {\bf 45} 5744

\bibitem{dagottoRev}
Dagotto E 1999 {\it Rep.\ Prog.\ Phys.\ } {\bf 62}  1525

\bibitem{uehara}
Uehara M, Nagata T, Akimitsu J, Takahashi H, M\^{o}ri N and Kinoshita K 1996
{\it J. Phys.\ Soc.\ Jpn.\ } {\bf 65} 2764

\bibitem{ishihara97}
Ishihara S, Yamanaka M and Nagaosa N 1997
{\it Phys.\ Rev.\ } B {\bf 56} 686

\bibitem{feiner}
Feiner L F, Oles A M and Zaanen J 1997
{\it Phys.\ Rev.\ Lett.\ } {\bf 78} 2799

\bibitem{khaliullin}
Khaliullin G and Maekawa S 2000 {\it Phys.\ Rev.\ Lett.\ } {\bf 85} 3950

\bibitem{khaliullin01}
Khaliullin G 2001 {\it Phys.\ Rev.\ } B {\bf 64} 212405

\bibitem{khaliullinyvo}
Khaliullin G, Horsch P and Ole\'{s} A M 2001
{\it Phys.\ Rev.\ Lett.\ } {\bf 86} 3879

\bibitem{sirker}
Sirker J and Khaliullin G 2003 {\it Phys.\ Rev.\ } B {\bf 67} 100408(R)

\bibitem{horschyvo}
Horsch P, Khaliullin G and Ole\'{s} A M 2003
{\it Phys.\ Rev.\ Lett.\ } {\bf 91} 257203

\bibitem{saitoh}
Saitoh E, Okamoto S, Takahashi K T, Tobe K, Yamamoto K, Kimura T, Ishihara S, Maekawa S and Tokura Y
2001 {\it Nature} {\bf 410} 180

\bibitem{ishihara97a}
Ishihara S, Inoue J and Maekawa S 1997 {\it Phys.\ Rev.\ } B {\bf 55} 8280

\bibitem{khaliullinokamoto02}
Khaliullin G and Okamoto S 2002
{\it Phys.\ Rev.\ Lett.\ } {\bf 89} 167201

\bibitem{khaliullinokamoto03}
Khaliullin G and Okamoto S 2003 {\it Phys.\ Rev.\ } B {\bf 68} 205109

\bibitem{ishiharaytio}
Ishihara S 2004 {\it Phys.\ Rev.\ } B {\bf 69} 075118

\bibitem{keimerlatio}
Keimer B, Casa D, Ivanov A, Lynn J W, von Zimmermann M, Hill J P, Gibbs D, Taguchi Y and Tokura Y
2000 {\it Phys.\ Rev.\ Lett.\ } {\bf 85} 3946

\bibitem{seidel03}
Seidel A, Marianetti C A, Chou F C, Ceder G and Lee P A 2003
{\it Phys.\ Rev.\ } B {\bf 67} 020405(R)

\bibitem{seidel04}
Seidel A and Lee P A 2004
{\it Phys.\ Rev.\ } B {\bf 69} 094419

\bibitem{ulrichyvo}
Ulrich C, Khaliullin G, Sirker J, Reehuis M, Ohl M,  Miyasaka S, Tokura Y and Keimer B
2003 {\it Phys.\ Rev.\ Lett.\ } {\bf 91} 257202

\bibitem{cwik}
Cwik M, Lorenz T, Baier J, M\"{u}ller R, Andr\'{e} G, Bour\'{e}e F, Lichtenberg F, Freimuth A,
Schmitz R, M\"{u}ller-Hartmann E and Braden M
2003 {\it Phys.\ Rev.\ } B {\bf 68} 060401(R)

\bibitem{cracolatio}
Craco L, Laad M S, Leoni S and M\"{u}ller-Hartmann E 2004
{\it Phys.\ Rev.\ } B {\bf 70} 195116
% cond-mat/0309370

\bibitem{haverkortlatio}
Haverkort M W, Hu Z, Tanaka A, Ghiringhelli G, Roth H, Cwik M, Lorenz T, Sch\"{u}{\ss}ler-Langeheine C,
Streltsov S V, Mylnikova A S, Anisimov V I, de Nadai C, Brookes N B, Hsieh H H,
Lin H-J, Chen C T, Mizokawa T, Taguchi Y, Tokura Y, Khomskii D I and Tjeng L H 2005
{\it Phys.\ Rev.\ Lett.\ } {\bf 94} 056401

\bibitem{mochizuki01}
Mochizuki M and Imada M 2001
{\it J. Phys.\ Soc.\ Jpn.\ } {\bf 70} 2872

\bibitem{mochizuki03}
Mochizuki M and Imada M 2003 {\it Phys.\ Rev.\ Lett.\ } {\bf 91} 167203

\bibitem{mochizuki03b}
Mochizuki M and Imada M 2004 {\it J. Phys.\ Soc.\ Jpn.\ } {\bf 73} 1833
% cond-mat/0312244

\bibitem{mochizuki04}
Mochizuki M and Imada M 2004 {\it New J. Phys.\ } {\bf 6} 154

\bibitem{fritschlatio}
Fritsch V, Hemberger J, Eremin M V, Krug von Nidda H-A, Lichtenberg F, Wehn R and Loidl A
2002 {\it Phys.\ Rev.\ } B {\bf 65} 212405

\bibitem{hemberger}
Hemberger J, Krug von Nidda H-A, Fritsch V, Deisenhofer J, Lobina S, Rudolf T, Lunkenheimer P,
Lichtenberg F, Loidl A, Bruns D and B\"{u}chner B 2003
{\it Phys.\ Rev.\ Lett.\ } {\bf 91} 066403

\bibitem{kiyama}
Kiyama T and Itoh M 2003 {\it Phys.\ Rev.\ Lett.\ } {\bf 91} 167202

\bibitem{harris03}
Harris A B, Yildirim T, Aharony A, Entin-Wohlmann O and Korenblit I Ya
2003 {\it Phys.\ Rev.\ Lett.\ } {\bf 91} 087206

\bibitem{harris04}
Harris A B, Yildirim T, Aharony A, Entin-Wohlmann O and Korenblit I Ya
2004 {\it Phys.\ Rev.\ } B {\bf 69} 035107

\bibitem{kikoin}
Kikoin K, Entin-Wohlmann O, Fleurov V and Aharony A 2003
{\it Phys.\ Rev.\ } B {\bf 67} 214418

\bibitem{pavarini}
Pavarini  E, Biermann S, Poteryaev A, Liechtenstein A I, Georges A and Andersen O K
2004 {\it Phys.\ Rev.\ Lett.\ } {\bf 92} 176403

\bibitem{solovyev04}
Solovyev I V 2004 {\it Phys.\ Rev.\ } B {\bf 69} 134403
%% cond-mat/0310581

\bibitem{nagaosa}
Fang Z and Nagaosa N 2004
{\it Phys.\ Rev.\ Lett.\ } {\bf 93} 176404

\bibitem{grueninger}
Gr\"{u}ninger M, R\"{u}ckamp R, Windt M and Freimuth A 2002 {\it Nature} {\bf 418} 39

\bibitem{krueger}
Kr\"{u}ger R, Schulz B, Naler S, Rauer R, Budelmann D, B\"{a}ckstr\"{o}m J, Kim K H, Cheong S-W,
Perebeinos V and R\"{u}bhausen M
2004 {\it Phys.\ Rev.\ Lett.\ } {\bf 92} 097203

\bibitem{martincarron}
Mart\'{\i}n-Carr\'{o}n L and de Andr\'{e}s A 2004
{\it Phys.\ Rev.\ Lett.\ } {\bf 92} 175501

\bibitem{ballhausen}
Ballhausen C F 1962
{\em Introduction to Ligand Field Theory} McGraw-Hill, New York

\bibitem{kikoinbook}
Kikoin K A and Fleurov V N 1994
{\it Transition Metal Impurities in Semiconductors. Electronic Structure and Physical Properties}
World Scientific, Singapore

\bibitem{nelson}
Nelson E D, Wong J Y and Schawlow A L 1967
{\it Phys.\ Rev.\ } {\bf 156} 298

\bibitem{schawlow}
Schawlow A L, Wood D L and Clogston A M 1959 {\it Phys.\ Rev.\ Lett.\ } {\bf 3} 271

\bibitem{sell}
Sell D D, Greene R L and White R M 1967 {\it Phys.\ Rev.\ } {\bf 158} 489

\bibitem{hasan}
Hasan M Z, Isaacs E D, Shen Z-X, Miller L L, Tsutsui K, Tohyama T and Maekawa S 2000
{\it Science} {\bf 288} 1811

\bibitem{khaliullinkilian}
Khaliullin G and Kilian R 2000
{\it Phys.\ Rev.\ } B {\bf 61} 3494

\bibitem{brink98}
van den Brink J, Stekelenburg W, Khomskii D I, Sawatzky G A, and Kugel K I
1998 {\it Phys.\ Rev.\ } B {\bf 58} 10276-10282

\bibitem{figgis}
Figgis B N and Hitchman M A 1999
{\em Ligand Field Theory and its Applications} Wiley

\bibitem{advances}
Gr\"{u}ninger M, Windt M, Benckiser E, Nunner T S, Schmidt K P, Uhrig  G S and Kopp T 2003
{\it Adv.\ Solid State Phys.\ } {\bf 43} 95

\bibitem{falck}
Falck J P, Perkins J D, Levy A, Kastner M A, Graybeal J M and Birgeneau R J 1994
{\it Phys.\ Rev.\ } B {\bf 49} 6246

\bibitem{dodge}
Schumacher A B, Dodge J S, Carnahan M A, Kaindl R A, Chemla D S and Miller LL 2001
{\it Phys.\ Rev.\ Lett.\ }{\bf 87} 127006

\bibitem{fromme}
Fromme B, M\"{o}ller M, Ansch\"{u}tz T, Bethke C Kisker E 1996
{\it Phys.\ Rev.\ Lett.\ } {\bf 77} 1548

\bibitem{tanabe}
Ferguson J, Guggenheim H J and Tanabe Y 1966
{\it J. Phys.\ Soc.\ Jpn.\ } {\bf 21} 692

\bibitem{kahn}
Cador O, Mathoni\`{e}re C and Kahn O 2000
{\it Inorg.\ Chem.\ } {\bf 39} 3799-3804

\bibitem{guillaume}
Guillaume M, Henggeler W, Furrer A, Eccleston R S and Trounov V 1995
{\it Phys.\ Rev.\ Lett.\ }{\bf 74} 3423

\bibitem{tanaka}
Tanaka Y, Baron A Q R, Kim Y-J, Thomas K J, Hill J P, Honda Z, Iga F,
Tsutsui S, Ishikawa D and Nelson C S 2004
{\it New J. Phys.\ }{\bf 6} 161

\bibitem{khaliullinpriv}
Khaliullin G {\it private communication}

\bibitem{losawa95}
Lorenzana J and Sawatzky G A 1995
{\it Phys.\ Rev.\ Lett.\ } {\bf 74} 1867

\bibitem{losawa95b}
Lorenzana J and Sawatzky G A 1995 {\it Phys. Rev.\ } B {\bf 52} 9576

\bibitem{suzuura}
Suzuura H, Yasuhara H, Furusaki A, Nagaosa N and Tokura Y 1996
{\it Phys.\ Rev.\ Lett.\ }{\bf 76} 2579

\bibitem{loeder}
Lorenzana J and Eder R 1997
{\it Phys.\ Rev.\ } B {\bf 55} R3358

\bibitem{perkins93}
Perkins J D, Graybeal J M, Kastner M A, Birgeneau R J, Falck J P and Greven M 1993
{\it Phys.\ Rev.\ Lett.\ }{\bf 71} 1621

\bibitem{perkins98}
Perkins J D, Birgeneau R J, Graybeal J M, Kastner M A and Kleinberg D S 1998
{\it Phys.\ Rev.\ } B {\bf 58} 9390

\bibitem{perkinsNi}
Perkins J D, Kleinberg D S, Kastner M A, Birgeneau R J, Endoh Y,
Yamada K and Hosoya S 1995
{\it Phys.\ Rev.\ } B {\bf 52} R9863

\bibitem{brink01}
van den Brink J 2001 {\it Phys.\ Rev.\ Lett.\ } {\bf 87} 217202

\bibitem{Cowan81}
Cowan R D 1981
\textit{The theory of atomic structure and spectra} University of California press, Berkely

\bibitem{SuganoTanabeKamimura70}
Sugano S, Tanabe Y and Kamimura H 1970
\textit{Multiplets of Transition-Metal Ions in Crystals}
Academic Press, New York and London

\bibitem{Harrison80}
Harrison W A 1980
\textit{Electronic Structure and the Properties of Solids}
W.H.Freeman and Company, San Francisco

\bibitem{Slater54}
Slater J C and Koster G F 1954
{\it Phys.\ Rev.\ } \textbf{94} 1498

\bibitem{Saitoh95}
Saitoh T, Bocquet A E, Mizokawa T and Fujimori A 1995
{\it Phys.\ Rev.\ } B \textbf{52} 5419

\bibitem{Tanaka94}
Tanaka A and Jo T 1994
{\it J. Phys.\ Soc.\ Jpn.\ } \textbf{63} 2788

\bibitem{kataevtiocl}
Kataev V, Baier J, M\"{o}ller A, Jongen L, Meyer G and Freimuth A 2003
{\it Phys.\ Rev.\ } B {\bf 68} 140405

\bibitem{lichtenberg}
Lichtenberg F, Widmer D, Bednorz J G, Williams T and Reller A 1991
{\it Z. Phys.\ } B {\bf 82} 211

\bibitem{geck}
Geck J, Wochner P, Kiele S, Klingeler R, Revcolevschi A, von Zimmermann M, and B\"{u}chner B
2004 {\it New J. Phys.\ } {\bf 6} 152

\bibitem{massarotti}
Massarotti V, Capsoni D, Bini M, Altomare A and Moliterni A G G 1999
{\it Zeitschrift f\"{u}r Kristallographie} {\bf 214} 205

\bibitem{sulewski}
Sulewski P E and Cheong S-W 1994
{\it Phys.\ Rev.\ } B {\bf 50} 551

\bibitem{sekar02}
Sekar C, Ruck K, Krabbes G, Teresiak A and Watanabe T
2002 {\it Physica} C {\bf 378} 678

\bibitem{deboer}
de Boer J J, Bright D and Helle J N 1972
{\it Acta Crystallographica} B {\bf 28} 3436

\bibitem{SNS}
Gr\"{u}ninger M, Windt M, Nunner T, Knetter C, Schmidt K P, Uhrig G S,
Kopp T, Freimuth A, Ammerahl U, B\"{u}chner B and Revcolevschi A 2002
{\it J. Phys.\ Chem.\ Solids } {\bf 63} 2167-2173

%%%%%%%%%%%%%%%%%%%%%% TiOCl

\bibitem{beynon}
Beynon R J and Wilson J A 1993
{\it J. Phys.: Condens. Matter} {\bf 5} 1983

\bibitem{imai}
Imai T and Chou F C 2003 {\it Preprint} cond-mat/0301425

\bibitem{lemmenstiocl}
Lemmens P, Choi K-Y, Caimi G, Degiorgi L, Kovaleva N N, Seidel A and Chou F C
2004 {\it Phys.\ Rev.\ } B {\bf 70} 134429
% cond-mat/0307502

\bibitem{caimi03}
Caimi G, Degiorgi L, Kovaleva N N, Lemmens P and Chou F C
2004 {\it Phys.\ Rev.\ } B {\bf 69} 125108
% cond-mat/0308273

\bibitem{caimi04}
Caimi G, Degiorgi L, Lemmens P and Chou F C 2004
{\it J. Phys.\ Cond.\ Mat.\ } {\bf 16} 5583
% cond-mat/0404502

\bibitem{sahadasgupta}
Saha-Dasgupta T, Valenti R, Rosner H and Gros C
2004 {\it Europhys.\ Lett.\ } {\bf 67} 63

\bibitem{hemberger05}
Hemberger J, Hoinkis M, Klemm M, Sing M, Claessen R, Horn S and Loidl A 2005
{\it Preprint} cond-mat/0501517

\bibitem{shaz}
Shaz  M, van Smaalen S, Palatinus L, Hoinkis M, Klemm M, Horn S and Claessen R
2005 {\it Phys.\ Rev.\ } B {\bf 71} 100405(R)

\bibitem{abel}
Abel E, Matan K, Chou F C and Lee Y S 2004
{\it Bull.\ Amer.\ Phys.\ Soc.\ } {\bf 49} 317
{\it (APS March Meeting Montreal 2004, Session D25.014)}

\bibitem{snigireva}
Snigireva E M, Troyanov S I and Rybakov V B 1990
{\it Zhurnal Neorganicheskoi Khimii} {\bf 35} 1945
(ICSD code 39314)

\bibitem{maule}
Maule C H, Tothill J N, Strange P and Wilson J A 1988
{\it J. Phys.\ C: Solid State Phys.\ } {\bf 21} 2153-2179

\bibitem{lemmenstiobr}
Lemmens P, Choi K-Y, Valenti R, Saha-Dasgupta T, Abel E, Lee Y S and Chou F C
2005 {\it Preprint} cond-mat/0501577

\bibitem{sasaki}
Sasaki T, Mizumaki M, Kato K, Watabe Y, Nishihata Y, Takata M and Akimitsu J
2005 {\it Preprint} cond-mat/0501691

\bibitem{rueckamp}
R\"{u}ckamp R, Baier J, Kriener M, Haverkort M W, Lorenz T, Uhrig G S,
Jongen L, M\"{o}ller A, Meyer G and Gr\"{u}ninger M 2005
{\it Preprint} cond-mat/0503409.

\bibitem{cracotiocl}
Craco L, Laad M and M\"{u}ller-Hartmann E
2004 {\it Preprint} cond-mat/0410472

\bibitem{sahadasgupta04b}
Saha-Dasgupta T, Lichtenstein A and Valenti R
2005 {\it Phys.\ Rev.\ } B {\bf 71} 153108

\bibitem{pisani}
Pisani L and Valenti R
2005 {\it Phys.\ Rev.\ } B {\bf 71} 180409

%%%%%%%%%%%%%%%%%%%%%%%%%%%%%%% RTiO3

\bibitem{barba}
Barba D, Jandl S, Nekvasil V, Mary\u{s}ko M, Divi\u{s} M,
Martin A A, Lin C T, Cardona M and Wolf T 2001
{\it Phys.\ Rev.\ } B {\bf 63} 054528

\bibitem{ulrichRaman}
Ulrich C, G\"{o}ssling A, Gr\"{u}ninger M, Guennou M, Roth H, Cwik M, Lorenz T, Khaliullin G
and Keimer B 2005 {\it Preprint} cond-mat/0503106

\bibitem{maclean}
MacLean D A, Ng H-N and Greedan J E 1979 {\em J. Sol.\ State Chem.\ } {\bf 30} 35

\bibitem{ulrichytio}
Ulrich C, Khaliullin G, Okamoto S, Reehuis M, Ivanov A, He H, Taguchi Y,
Tokura Y and Keimer B 2002
{\it Phys.\ Rev.\ Lett.\ } {\bf 89} 167202

\bibitem{ytiostructure}
Zubkov V G, Berger I F, Artamonova A M and Bazuyev G V 1984
{\it Kristallografiya} {\bf 29} 494-497

\bibitem{schmitz}
Schmitz R, Entin-Wohlman O, Aharony A, Harris A B and M\"{u}ller-Hartmann E
2004 {\it Preprint} cond-mat/0411583

\bibitem{schmitzJ}
Schmitz R, Entin-Wohlman O, Aharony A, Harris A B and M\"{u}ller-Hartmann E
2005  {\it Phys.\ Rev.\ } B {\bf 71} 144412

\bibitem{held}
Held K, Ulmke M, Bl\"{u}mer N and Vollhardt D 1997
{\it Phys.\ Rev.\ } B {\bf 56} 14469

%%%%%%% LaMnO3

\bibitem{brinkkhomskii}
van den Brink J, Khaliullin G and Khomskii D I 2002
{\it Preprint} cond-mat/0206053

\bibitem{murakami}
Murakami Y, Hill J P, Gibbs D, Blume M, Koyama I, Tanaka M, Kawata H, Arima T, Tokura Y,
Hirota K and Endoh Y 1998
{\it Phys.\ Rev.\ Lett.\ }{\bf 81} 582

\bibitem{rodriguez}
Rodriguez-Carvajal J, Hennion M, Moussa F, Moudden A H,
Pinsard L, and Revcolevschi A 1998 {\it Phys.\ Rev.\ } B {\bf 57} R3189

\bibitem{chatterji}
Chatterji T, Fauth F, Ouladdiaf B, Mandal P and Ghosh B 2003
{\it Phys.\ Rev.\ } B {\bf 68} 052406

\bibitem{brink04}
van den Brink J 2004
{\it New J. Phys.\ } {\bf 6} 201

\bibitem{braden}
Reichardt W and Braden M 1999 {\it Physica } b {\bf 263-264} 416

\bibitem{lemmenslamno}
Choi K Y, Lemmens P, G\"{u}ntherodt G, Pashkevich Y G, Gnezdilov V P, Reutler P,
Pinsard-Gaudart L, B\"{u}chner B, and Revcolevschi A
2005 {\em Preprint} cond-mat/0503460.

\bibitem{saitohreply}
Saitoh E, Okamoto S, Tobe K, Yamamoto K, Kimura T, Ishihara S,
Maekawa S, and Tokura Y 2002 {\it Nature} {\bf 418} 40

\bibitem{paolone}
Paolone A, Roy P, Pimenov A, Loidl A, Mel'nikov O K, and Shapiro A Y
2000 {\it Phys.\ Rev.\ } B {\bf 61} 11255

\bibitem{perebeinos01}
Perebeinos V and Allen P B 2001 {\it Phys.\ Rev.\ } B {\bf 64} 085118

\bibitem{reichardt}
Reichardt W and Braden M {\it private communictaion}

\bibitem{grueninger99}
Gr\"{u}ninger M, van der Marel D, Damascelli A, Zibold A, Geserich H P,
Erb A, Kl\"{a}ser M, Wolf T, Nunner T, and Kopp T 1999
{\it Physica} C {\bf 317-318} 286-291

\bibitem{chaplot}
Chaplot S L, Reichardt W, Pintschovius L, and Pyka N 1995
{\it Phys.\ Rev.\ } B {\bf 52} 7230

\bibitem{tobe}
Tobe K, Kimura T, Okimoto Y and Tokura Y 2001
{\it Phys.\ Rev.\ } B {\bf 64} 184421

\bibitem{kovaleva}
Kovaleva N N, Boris A V, Bernhard C, Kulakov A, Pimenov A, Balbashov A M,
Khaliullin G and Keimer B 2004
{\it Phys.\ Rev.\ Lett.\ } {\bf 93} 147204
% cond-mat/0405509

\bibitem{perebeinos}
Allen P B and Perebeinos V 1999
{\it Phys.\ Rev.\ Lett.\ } {\bf 83} 4828

\bibitem{millis}
Millis A J 1996
{\it Phys.\ Rev.\ } B {\bf 53} 8434

\bibitem{pickett}
Pickett W E and Singh D J 1996
{\it Phys.\ Rev.\ } B {\bf 53} 1146

\bibitem{solovyev}
Solovyev I, Hamada N and Terakura K 1996
{\it Phys.\ Rev.\ } B {\bf 53} 7158

\bibitem{elfimov}
Elfimov I S, Anisimov V I and Sawatzky G A 1999
{\it Phys.\ Rev.\ Lett.\ }{\bf 82} 4264

\bibitem{ahnmillis}
Ahn K H and Millis A J 2000
{\it Phys.\ Rev.\ } B {\bf 61} 13545

\bibitem{bala}
Bala J and Oles A M 2000
{\it Phys.\ Rev.\ } B {\bf 62} R6085

%%%%%%%%%%% Y2BaNiO5

\bibitem{darriet}
Darriet J and Regnault L P 1993 {\it Solid State Comm.\ } {\bf 86} 409

\bibitem{xu}
Xu G, Aeppli G,Bisher M E, Broholm C, DiTusa J F, Frost C D, Ito T, Oka K , Paul R L,
Takagi H and Treacy M M J 2000
{\it Science} {\bf 289} 419

\bibitem{nunnerpriv}
Nunner T and Kopp T {\it private communication}

%%%%%%%% cuprates

\bibitem{grant92}
Grant J B and McMahan A K 1992
{\it Phys.\ Rev.\ } B {\bf 46} 8440

\bibitem{martin93}
Martin R L and Hay P J 1993
{\it J. Chem.\ Phys.\ }{\bf 98} 8680

\bibitem{eskes90}
Eskes H, Tjeng L H and Sawatzky G A 1990
{\it Phys.\ Rev.\ } B {\bf 41} 288

\bibitem{graafthesis}
de Graaf C 1998
{\it Ph.D. Thesis} University of Groningen, The Netherlands.

\bibitem{geserich88}
Geserich H P, Scheiber G, Geerk J, Li H C, Linker G, Assmus W and Weber W 1988
{\it Europhys.\ Lett.\ }{\bf 6} 277

\bibitem{kuiper}
Kuiper P, Guo J-H, S{\aa}the C, Duda L-C, Nordgren J, Pothuizen J J M, de Groot F M F
and Sawatzky G A 1998 {\it Phys.\ Rev.\ Lett.\ }{\bf 80} 5204

\bibitem{teske}
Teske C L and M\"{u}ller-Buschbaum H K 1969
{\it Zeitschrift f\"{u}r Anorganische und Allgemeine Chemie} {\bf 370} 134

\bibitem{kimcacu2o3}
Kim T K, Rosner H, Drechsler S-L, Hu Z, Sekar C, Krabbes G, M\'{a}lek J, Knupfer M, Fink J and Eschrig H
2003 {\it Phys.\ Rev.\ } B {\bf 67} 024516

\bibitem{yamashita}
Yamashita Y and Ueda K 2000
{\it Phys.\ Rev.\ Lett.\ }{\bf 85} 4960

\bibitem{kotovtetra}
Kotov V N, Zhitomirsky M E and Sushkov O P 2001
{\it Phys.\ Rev.\ } B {\bf 63} 064412

\bibitem{lemmenscu2te2}
Lemmens P, Choi K-Y, Kaul E E, Geibel C, Becker K, Brenig W, Valenti R, Gros C,
Johnsson M, Millet P and Mila F 2001
{\it Phys.\ Rev.\ Lett.\ }{\bf 87} 227201

\bibitem{brenigcu2te2}
Brenig W 2003 {\it Phys.\ Rev.\ } B {\bf 67} 064402

\endbib

\end{document}